\documentclass{article}

\PassOptionsToPackage{numbers, compress}{natbib}
\usepackage[preprint]{neurips_2026}

\usepackage[utf8]{inputenc} 
\usepackage[T1]{fontenc}    
\usepackage{hyperref}       
\usepackage{url}            
\usepackage{booktabs}       
\usepackage{array}          
\usepackage{longtable}      
\usepackage{float}          
\usepackage{amsfonts}       
\usepackage{nicefrac}       
\usepackage{microtype}      
\usepackage{xcolor}         

\usepackage{graphicx}
\usepackage{amsmath}

\title{Sympatheia: Emotionally Adaptive Voice Assistant with Continuous Affect Conditioning}

\author{%
  Sukru Samet~Dindar\\
  Department of Electrical Engineering\\
  Columbia University\\
  New York, NY 10025 \\
  \texttt{sd3705@columbia.edu} \\
  \And
  Riki~Shimizu\\
  Department of Electrical Engineering\\
  Columbia University\\
  New York, NY 10025 \\
  \texttt{rs4613@columbia.edu} \\
  \And
  Xilin~Jiang\\
  Department of Electrical Engineering\\
  Columbia University\\
  New York, NY 10025 \\
  \texttt{xj2289@columbia.edu} \\
  \And
  Nima~Mesgarani\\
  Department of Electrical Engineering\\
  Columbia University\\
  New York, NY 10025 \\
  \texttt{nm2764@columbia.edu} \\
}

\begin{document}

\maketitle

\begin{abstract}
  Empathetic spoken dialogue systems must infer a user's emotional state to respond appropriately, yet everyday speech often carries weak, neutral, or ambiguous affective cues. To address this, we introduce \textsc{Sympatheia}, a speech-to-speech dialogue framework conditioned on affect inferred from the user's speech and, when available, explicit affect specifications provided as a continuous valence--arousal (VA) control signal by a multimodal sensing module or user interface. To train our model, we construct \textsc{Sympatheia}-18k, an emotion-conditioned synthetic spoken dialogue corpus with 12 emotion anchors. This dataset includes an emotional split for learning affective speech behavior, and a neutral split that pairs emotionally neutral queries with multiple emotion-conditioned responses to isolate explicit emotion control in emotionally ambiguous cases. Empirical results show that \textsc{Sympatheia} outperforms speech conversational baselines in generating responses whose semantic content and spoken delivery are both emotionally appropriate. We further show that the same VA interface can integrate emotion estimates from diverse sensing modules, including facial expression, biosignals, and textual affect descriptions, improving response alignment when speech alone provides limited emotional evidence. These results suggest that continuous affect conditioning is an effective practical step for building emotionally adaptive voice assistants.
\end{abstract}

\section{Introduction}

\begin{figure}[t]
  \centering
  \includegraphics[width=0.85\linewidth]{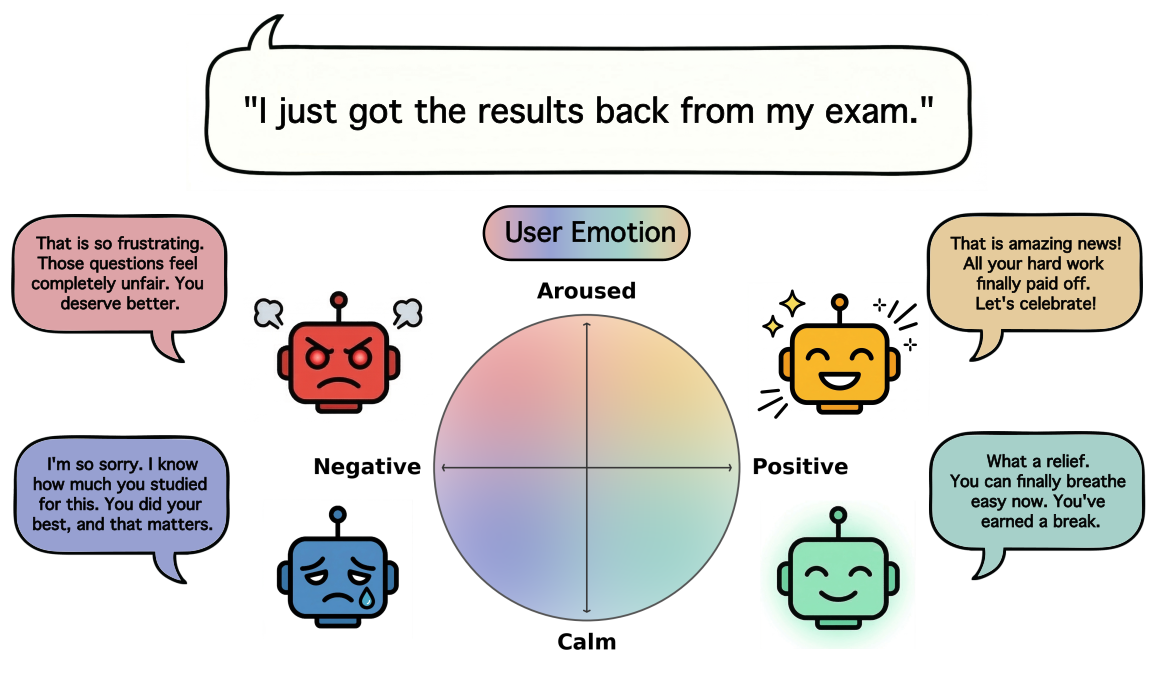}
  \caption{Motivation for \textsc{Sympatheia}. The same user utterance may call for different responses depending on the speaker’s emotional state, illustrated in a valence-arousal space. The system aims to generate spoken responses that are both semantically appropriate and emotionally aligned.}
  \label{fig:intro-overview}
\end{figure}

Voice-based conversational agents are becoming increasingly capable of flexible and open-ended conversational interaction, driven by recent advances in large language models (LLMs) \citep{brown2020language,ouyang2022training}. As these systems evolve from task-oriented tools to companions for extended interaction, response quality depends not only on factual correctness but also on interpersonal sensitivity. Emotional alignment plays an important role in effective communication, as it can make the assistant feel more natural and human-like, reduce misunderstanding, and support interactions that feel more comforting and engaging. In contrast, affectively mismatched responses can alter how a message is perceived and weaken the overall quality of the interaction. Figure~\ref{fig:intro-overview} illustrates why this capability is essential for an empathetic spoken companion.

Prior work on empathetic and emotion-aware dialogue has made important progress in generating more contextually appropriate, supportive, and affect-sensitive responses \citep{rashkin2019towards}, but two key gaps remain. First, many methods are optimized for spoken inputs with explicit, high-intensity emotional cues. In practice, however, a central challenge is that everyday speech often contains subtle, ambiguous, or mixed affective states that may not be reliably inferred from lexical or acoustic cues alone \citep{scherer2003vocal,zeng2009survey}. Human listeners naturally resolve this ambiguity by integrating multiple cues and adapting their responses empathetically. In contrast, current spoken dialogue systems often have limited mechanisms for combining ambiguous emotional cues in user's speech with external emotional cues beyond speech. As a result, they may generate semantically appropriate yet emotionally misaligned responses.

Second, emotion is often represented using a fixed set of discrete categories, which can be overly restrictive for natural interaction. Evidence from affective computing suggests that real-world emotional expression varies across speakers and cultures, and that continuous representations such as valence-arousal can better capture this nuance \citep{russell1980circumplex}. However, many widely used emotion resources are still organized around a small set of discrete emotion categories, which can limit how well they capture the variability and subtlety of naturalistic speech \citep{busso2008iemocap,poria2019meld}.

To address these limitations, we introduce \textsc{Sympatheia}, an emotion-conditioned voice dialogue model for adaptive empathetic response generation. \textsc{Sympatheia} combines (i) implicit affect inference from user speech with (ii) optional explicit affect control through a continuous valence--arousal (VA) interface. During training, the 12 named emotions serve as interpretable anchor points in the VA plane. Each anchor is encoded by a valence--arousal coordinate, allowing the model to learn affect control in a continuous space rather than as a closed set of class labels. At inference time, this same interface can accept continuous VA coordinates, discrete labels mapped to anchors, or estimates from external affect modules such as biosignals, facial expressions, and textual affect descriptions.\footnote{Code and audio demos are available at \url{https://github.com/susameddin/sympatheia}.}

Our main contributions are: (1) We propose \textsc{Sympatheia}, an empathetic voice chatbot framework conditioned on both speech-inferred affect and explicit valence--arousal cues from emotion sensing modules or a user interface. Across different evaluation settings, \textsc{Sympatheia} outperforms strong spoken-dialogue baselines in empathy-oriented response quality. (2) We introduce the \textsc{Sympatheia-18k} dataset with 18k emotion-conditioned synthetic spoken query--response pairs, including both emotionally expressive and neutral interactions. (3) We present a modular interface for incorporating external affect sources, including facial expressions, biosignals such as EEG, ECG or eye tracking, and textual affect descriptions, enabling multimodal emotion grounding in real-world deployments.

\section{Related Work}

\textbf{Text-based empathetic dialogue.} Empathetic dialogue research has been driven largely by text-based benchmarks and generators. EmpatheticDialogues established the standard benchmark for supportive open-domain response generation \citep{rashkin2019towards}, and subsequent models such as MoEL and MIME improved text-based empathy by explicitly routing generation through inferred listener types or emotion mixtures \citep{lin2019moel,majumder2020mime}. These systems established the task formulation most later work inherits, but they primarily reason over lexical context and discrete emotion supervision rather than prosody or other paralinguistic cues.

\textbf{Emotion datasets and affect representation.} Frequently used emotion datasets span both categorical and continuous annotation schemes. IEMOCAP and MELD have supported emotion recognition in dialogue with discrete labels and multimodal context \citep{busso2008iemocap,poria2019meld}, while resources such as MSP-Podcast, AffectNet, and DEAP include dimensional affect annotations such as valence and arousal across speech, face, and physiological modalities \citep{lotfian2019building,busso2025msppodcast,mollahosseini2017affectnet,koelstra2012deap}. This line of work motivates valence--arousal as a compact interface for external sensing modules, because it can represent graded, mixed, and speaker-dependent affect more flexibly than a small fixed taxonomy \citep{russell1980circumplex}.

\textbf{Speech-native dialogue models.} Recent spoken-dialogue research has moved beyond cascaded automatic speech recognition (ASR) $\rightarrow$ LLM $\rightarrow$ text-to-speech (TTS) pipelines toward models that consume and/or generate speech more directly. Early speech-language systems such as SpeechGPT explored cross-modal conversational modeling \citep{zhang2023speechgpt}, and newer systems including Moshi, GLM-4-Voice, Qwen3-Omni, and Kimi-Audio further emphasize real-time interaction and preservation of timing or other paralinguistic cues \citep{defossez2024moshi,zeng2024glm4voice,xu2025qwen3omni,kimiteam2025kimiaudio}. These models advance speech-native interaction, but they do not primarily formulate empathy as a controllable response-generation objective.

\textbf{Empathetic spoken dialogue.} Closest to our setting are recent empathetic speech-language systems. BLSP-Emo aligns spoken inputs with empathetic text continuations, while OpenS2S, Empathy Omni, and OSUM-EChat extend empathetic dialogue to speech response generation \citep{wang2024blspemo,wang2025opens2s,wang2025empathyomni,geng2025osumechat}. These systems primarily learn affective or paralinguistic context from the user's spoken query itself. \textsc{Sympatheia} preserves this speech-based affect inference, but additionally exposes emotion as an optional continuous valence--arousal control signal, allowing external sensors or user controls to guide response generation when speech alone is neutral, subtle, or ambiguous.

\section{Methodology}

\begin{figure}[t]
  \centering
  \includegraphics[width=0.95\linewidth]{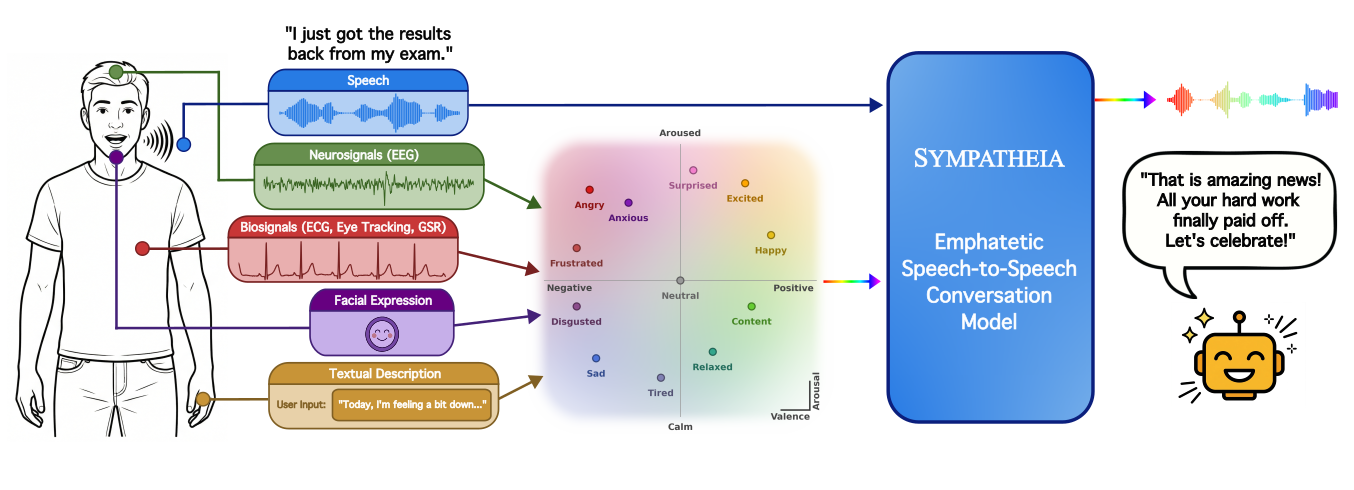}
  \caption{System overview of \textsc{Sympatheia}. The core speech-to-speech dialogue model uses affective cues in the user's spoken query directly and can additionally receive explicit valence--arousal conditioning from pluggable emotion sensing modules, such as facial expression, EEG/physiological signals, text descriptions, or direct interface selection. These optional external estimates are converted into a shared VA interface that guides emotionally aligned spoken response generation.}
  \label{fig:sympatheia-architecture}
\end{figure}

\subsection{System overview}
\label{sec:system-overview}
\textsc{Sympatheia} is an empathetic speech-to-speech conversational model, designed to generate spoken responses that are both semantically appropriate and emotionally aligned. Figure~\ref{fig:sympatheia-architecture} shows the system overview of \textsc{Sympatheia}. While the model already infers user affect from the semantic and acoustic cues present in the spoken query, it also supports system or user-provided external affect information when available. In the broader deployment setting, the overall \textsc{Sympatheia} system combines this core dialogue model with optional pluggable emotion sensing modules, yielding two components: (i) the \textsc{Sympatheia} speech-to-speech model itself and (ii) upstream emotion recognition modules that provide additional affect estimates from other available modalities, such as facial expressions, biosignals such as EEG and ECG, textual affect descriptions or affect selection in user interface. These external signals can provide useful complementary information when the user's emotional state is subtle, ambiguous, or only partially expressed in speech.

The interface between these components is intentionally minimal: \textsc{Sympatheia} accepts an optional continuous valence--arousal (VA) pair $z=(v,a)$, where $v\in[-1,1]$ denotes affective polarity and $a\in[-1,1]$ denotes activation. VA representations are widely used in emotion sensing, making them a practical shared interface for incorporating external affect estimates. Compared with discrete emotion labels alone, continuous VA coordinates better capture the graded, mixed, and subjective nature of real-world affect, while enabling interpolation between affective states and explicit control over emotional intensity. This design decouples sensing from generation, allowing \textsc{Sympatheia} to integrate different emotion recognition systems without architecture-specific engineering or retraining the conversational backbone. While the model can also consume discrete emotion labels, continuous conditioning provides greater flexibility for adaptive and personalized response generation.

\begin{figure}[t]
  \centering
  \includegraphics[width=0.65\linewidth]{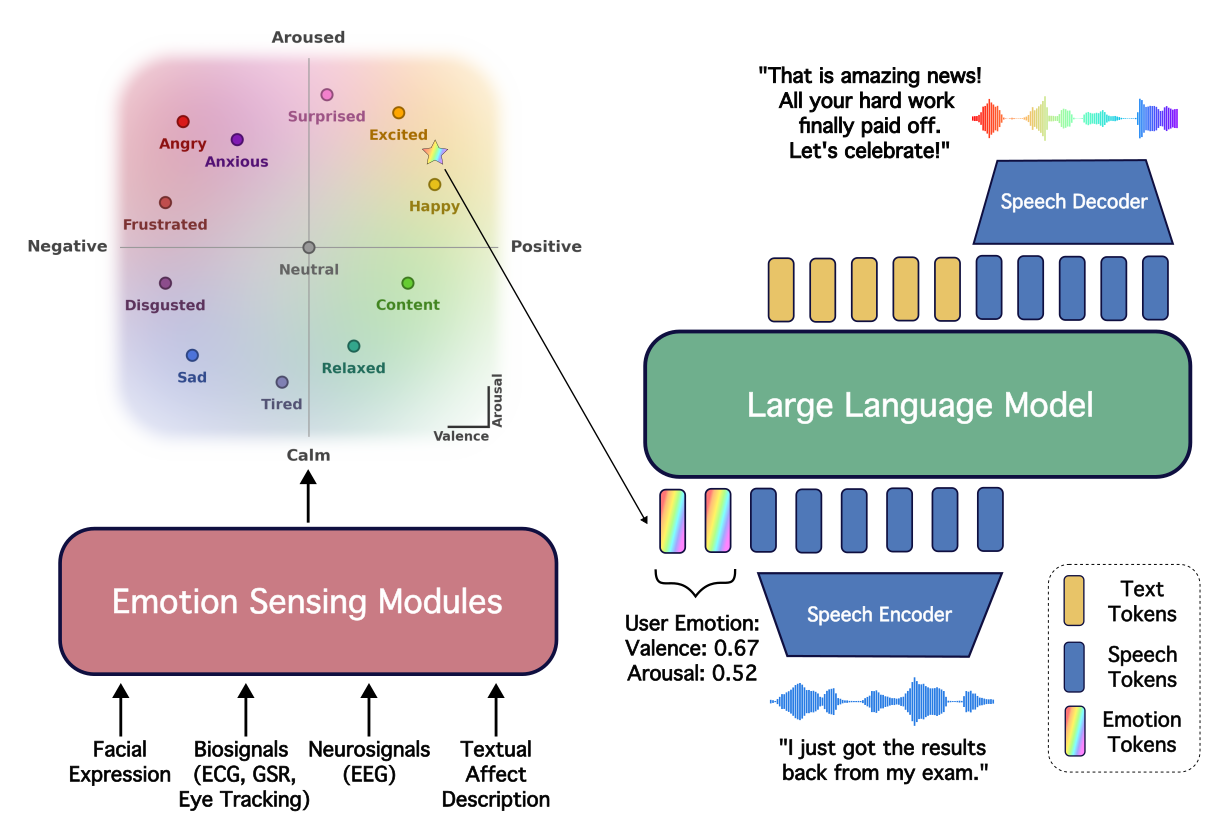}
  \caption{The architecture of the \textsc{Sympatheia} speech-to-speech model. Input audio is encoded into discrete speech tokens, while optional emotion recognition modules map facial expression, EEG/physiological signals, or user-provided affect descriptions into continuous valence--arousal (VA) values. The VA values and user speech tokens are fed to the language model, which generates affect-conditioned response speech tokens that the speech decoder renders as waveform audio.}
  \label{fig:architecture}
\end{figure}

\subsection{Emotion-conditioned speech generation}
\label{sec:emotion-conditioned-generation}
The architecture of \textsc{Sympatheia} is shown in Figure~\ref{fig:architecture}. Our speech-to-speech backbone follows GLM-4-Voice, an end-to-end speech-language model for spoken dialogue \citep{zeng2024glm4voice}. The model has three main components. First, a WhisperVQ speech tokenizer converts input audio into ultra-low-bitrate, single-codebook discrete tokens at 12.5 Hz using a vector-quantized bottleneck in a Whisper encoder \citep{zeng2024glm4voice,radford2022whisper}. Second, GLM-4-Voice-9B, initialized from GLM-4-9B and aligned for the speech modality, autoregressively processes the input sequence and generates response speech tokens. Finally, a streaming flow-matching speech decoder converts the generated tokens as waveform audio.

Although the base model supports coherent spoken interaction, it is not explicitly optimized to produce emotionally appropriate responses (as shown in Table~\ref{tab:overall-results}). We therefore expose affect as a controllable conditioning variable by inserting a continuous valence--arousal (VA) pair into the system prompt: \texttt{User emotion (valence=}$v$\texttt{, arousal=}$a$\texttt{).} To make affect conditioning reliable, we fine-tune the GLM-4-Voice-9B language model on \textsc{Sympatheia}-18k (Section~\ref{sec:dataset}) using Low-Rank Adaptation (LoRA) \citep{hu2022lora}, allowing the model to learn affect control while preserving its general conversational ability. During training, ground-truth VA values are provided in the system prompt, and the model learns to associate each region of the affect space with appropriate response content and vocal delivery. At inference time, the same VA fields can be supplied by any upstream emotion sensing module when external estimates are available; otherwise the VA prompt is omitted entirely.

We train \textsc{Sympatheia} with 12 emotion anchors in the VA plane: \emph{happy}, \emph{sad}, \emph{angry}, \emph{excited}, \emph{frustrated}, \emph{anxious}, \emph{relaxed}, \emph{surprised}, \emph{disgusted}, \emph{tired}, \emph{content}, and \emph{neutral}. Although these anchors correspond to discrete emotion names in the dataset, each anchor is represented by a coordinate in the continuous VA plane. This supervision reflects the geometry of affect space: nearby anchors correspond to related affective states, and intermediate coordinates can represent mixed or graded affect. This encourages the model to learn a continuous affect representation rather than a set of isolated labels, allowing VA values beyond the training anchors to be used as meaningful conditioning signals at inference time. The anchors span low/high arousal and negative/positive valence regions, as shown in Figure~\ref{fig:architecture}. The exact anchor values and the coordinate-selection rationale are provided in Appendix~\ref{app:va-anchor-coordinates}.

To make the model robust to missing or noisy affect estimates, we drop the VA condition randomly for one third of the non-neutral training samples. When the condition is absent, the model operates directly on the input speech tokens as a regular speech conversation model. When the condition is present, the model learns to use the VA pair as an explicit but optional control signal rather than treating it as a required input. This dropout encourages the model to infer affective context from the speech input itself using implicit semantic and acoustic cues when no external cue is available.

\subsection{Dataset creation}
\label{sec:dataset}

A key bottleneck for empathetic speech dialogue systems is the lack of widely adopted datasets for emotion-conditioned speech-to-speech interaction, especially datasets expressing the same intent under multiple affective states. Existing resources for empathetic dialogue are often text-only \citep{rashkin2019towards,lin2019moel,majumder2020mime} and therefore miss vocal emotional signals such as prosody, pacing, and intensity. Recent speech-native data construction efforts synthesize empathetic spoken dialogues \citep{wang2025opens2s}, but do not systematically pair the same query with different emotion-conditioned responses, a contrastive structure that can help models distinguish between affective states more clearly. We address these gaps by constructing a synthetic corpus for controllable emotional spoken dialogue.\footnote{The full \textsc{Sympatheia}-18k dataset is available at \url{https://huggingface.co/datasets/susameddin/Sympatheia-18k}. Examples are available at demo: \url{https://susameddin.github.io/sympatheia/}.}

\textsc{Sympatheia}-18k consists of two complementary splits. The \textbf{Emotional} split contains roughly 12k examples across the 12 target emotions, with about 1k examples per emotion, pairing affect-rich user queries with emotion-appropriate spoken responses to teach semantic and acoustic emotional alignment. The \textbf{Neutral} split is designed to isolate explicit affect control: we generate 500 emotionally neutral user queries and pair each query with 12 assistant responses, one for each target emotion. As the user speech provides little direct affective evidence, the desired response style must be determined primarily from the VA condition. This produces 6k examples that teach the model to use externally provided affect signals when speech is neutral, subtle, or ambiguous. Together, these splits expose the model to both emotionally expressive speech and near-neutral speech with explicit affect conditioning.

To generate the dataset, we first use Qwen3-32B \citep{yang2025qwen3} to generate query--response pairs for each target emotion under emotion-specific empathetic response strategies rather than simple emotion mirroring. The goal is to produce psychologically supportive responses for the user's affective state, such as gentle reassurance for sad users or grounded validation for angry users. We then synthesize emotionally styled audio from the generated text using Qwen3-TTS \citep{hu2026qwen3tts}, which provides emotion-consistent tone, prosody, and pacing. Finally, each sample is paired with VA metadata in the system prompt for training. Text generation establishes emotionally appropriate response semantics, while TTS supplies the acoustic realization needed for speech-to-speech fine-tuning. Dataset-generation prompts, response strategies, and TTS styles are provided in Appendices~\ref{app:dataset-creation-prompts} and~\ref{app:emotion-strategies}.

\subsection{Multimodal emotion sensing modules}
\label{sec:multimodal-sensing}
\textsc{Sympatheia} is designed to be modality-agnostic: beyond implicit speech emotion recognition, each available external modality is handled by an emotion recognition module whose output is converted into the same continuous valence--arousal (VA) interface. Each classifier produces a softmax distribution over its native emotion taxonomy, and the adaptation layer maps these probabilities to the canonical \textsc{Sympatheia} VA anchors by computing an expected affect coordinate:
\begin{equation}
\hat{z}_m = \sum_{k=1}^{K_m} p_m(y_k \mid x_m)\, \mu(y_k), \qquad \hat{z}_m \in [-1,1]^2,
\end{equation}
where $p_m(y_k \mid x_m)$ is the probability assigned by modality $m$ to class $y_k$, and $\mu(y_k)$ is the corresponding anchor coordinate. The probability-weighted mapping is useful behaviorally: it preserves uncertainty, interpolates between nearby emotions, and modulates emotional intensity instead of forcing inherently continuous affective states into a single discrete label. The shared VA plane provides a common representation across datasets whose emotion labels and implicit affect conventions do not perfectly match. We group external affect sensing into three modules. Additional modality-level details are provided in Appendix~\ref{app:sensing-details}.

\textbf{Facial expression.} Facial expressions provide a direct visual cue about affective state. We evaluate this modality on the AffectNet+ validation set, which provides facial expression annotations for in-the-wild face images \citep{mollahosseini2017affectnet}. We adopt an EfficientNet-style facial expression classifier following HSEmotion \citep{savchenko2022hsemotion}. To test whether this pathway works beyond offline simulations, we also conduct an end-to-end human facial expression recording study with 10 subjects, who act prompted emotions while speaking neutral-content queries. Their webcam video is processed into VA estimates and their recorded speech is used directly as the speech-to-speech model input. Study details are provided in Appendix~\ref{app:human-face-recording-study}, and project code includes a real-time demo implementation of the same pipeline.

\textbf{Biosignals.} Biosignals capture affect through nonverbal physiological and behavioral responses, including neuroelectric activity (EEG), eye tracking, cardiac activity (ECG), and skin conductance (GSR). For EEG and eye tracking, we use SEED-VII, a multimodal emotion dataset with synchronized neural and ocular recordings \citep{jiang2025seedvii}, and adopt the Multi-view Adaptive Emotion Transformer from the same work. For peripheral physiology, we use YAAD, which contains wearable ECG and GSR recordings for affect recognition \citep{dar2022yaad}, and train one-dimensional residual-network classifiers \citep{he2016resnet}.

\textbf{Textual affect description.} Textual affect descriptions support cases where a user explicitly states their emotional state rather than relying on sensed signals. We evaluate this modality on ISEAR, a dataset of self-reported emotional experiences \citep{scherer1994evidence}. We adopt a DistilRoBERTa-based classifier \citep{hartmann2022emotionenglish}.

\section{Results}

\subsection{Empathetic Response Evaluation}
\label{sec:empathetic-response}

\begin{table}[t]
  \centering
  \caption{Speech-generation empathy evaluation across \textsc{Sympatheia} and VoiceBench settings. Higher is better for empathy scores and Emotion MOS; lower is better for semantic and lexical similarity as these metrics measure response invariance across different target emotions.}
  \label{tab:overall-results}
  \begingroup
  \setlength{\tabcolsep}{3pt}
  \begin{tabular}{@{}lcccccc@{}}
    \toprule
    Model & \shortstack{Sympatheia-\\Neutral $\uparrow$} & \shortstack{Sympatheia-\\Emotional $\uparrow$} & \shortstack{VoiceBench-\\CommonEval $\uparrow$} & \shortstack{Emotion \\MOS $\uparrow$} & \shortstack{Semantic\\Similarity $\downarrow$} & \shortstack{Lexical\\Similarity $\downarrow$} \\
    \midrule
    \textsc{Sympatheia} & \textbf{4.37} & \textbf{4.74} & \textbf{4.22} & \textbf{3.86} & \textbf{0.801} & \textbf{0.223} \\
    GLM-4-Voice & 1.76 & 3.80 & 1.51 & 2.23 & 0.866 & 0.459 \\
    Qwen3-Omni & 2.59 & 4.69 & 1.88 & 3.32 & 0.857 & 0.397 \\
    Qwen2.5-Omni & 1.75 & 3.53 & 1.54 & 2.56 & 0.919 & 0.650 \\
    Kimi-Audio & 3.64 & 4.03 & 3.75 & 2.95 & 0.835 & 0.381 \\
    OpenS2S & 2.34 & 4.08 & 1.55 & 2.42 & 0.863 & 0.441 \\
    OSUM-EChat & 1.77 & 3.93 & 2.03 & 2.18 & 0.844 & 0.391 \\
    \bottomrule
  \end{tabular}
  \endgroup
\end{table}

Evaluating empathetic spoken dialogue is challenging because response quality depends on affective appropriateness, contextually supportive content, and how emotion is expressed in speech. We therefore use complementary evaluations: an audio-capable LLM-as-a-judge protocol for scalable system comparison, and a human Emotion MOS study for direct assessment of perceived emotional appropriateness. For the automated evaluation, following recent spoken-dialogue judge protocols \citep{yan2025urobench}, we use Qwen3-Omni 30B so that prosody, pacing, and vocal affect are evaluated alongside content rather than lost in transcription-only scoring. Detailed judge prompts are provided in Appendices~\ref{app:judge-prompts}.

Table~\ref{tab:overall-results} reports automated empathy scores on three settings: Sympatheia-Neutral, where the user speech is neutral but the system prompt supplies the target emotion; Sympatheia-Emotional, where no system receives an external emotion prompt and affect must be inferred from the expressive user speech; and VoiceBench-CommonEval, which uses real neutral spoken queries from VoiceBench \citep{chen2024voicebench}. In Sympatheia-Neutral and VoiceBench-CommonEval, \textsc{Sympatheia} receives continuous VA values, while baselines receive a discrete emotion label in the system prompt (e.g. "The user is angry."). Across all settings, \textsc{Sympatheia} achieves the highest empathy scores, with the largest margin when explicit affect conditioning is most important. Example generations and per-emotion results for \textsc{Sympatheia} are provided in Appendix Tables~\ref{tab:qualitative-examples} and~\ref{tab:sympatheia-per-emotion-results}, respectively.

We conduct a human Emotion MOS evaluation to directly assess whether generated responses are emotionally appropriate under explicit affect conditioning. We randomly select neutral queries and ask human evaluators to judge whether each response adapts appropriately to the stated user emotion. For each query, responses from the seven systems are shown in randomized order, and evaluators rate each response on a 1--5 scale. The study includes 20 participants recruited through Prolific \citep{palan2018prolific}; each participant rates all seven system responses for six queries to keep the annotation time reasonable. The 12 target emotions are distributed uniformly across the study, and the reported Emotion MOS is the mean score across ratings. As shown in Table~\ref{tab:overall-results}, \textsc{Sympatheia} achieves the highest Emotion MOS value across all models. Detailed protocol information is provided in Appendix~\ref{app:emotion-mos-evaluation}.

We also evaluate whether models actually adjust their response when only the stated user emotion changes. A common failure mode for baselines is to produce the same response, or only a minimally edited variant, for different emotion prompts. For each neutral query, we therefore compute pairwise similarity between responses generated under different target emotions, using BERTScore F1 for semantic similarity \citep{zhang2020bertscore} and ROUGE-L for lexical similarity \citep{lin2004rouge}, then average over all emotion pairs. Because these metrics measure response invariance, lower values suggest that a model varies its responses more across target emotions. Interpreted together with the empathy and MOS scores, \textsc{Sympatheia}'s lower semantic and lexical similarity provides additional evidence that it adjusts response content according to the user's stated affect rather than relying on generic templates.

\begin{table}[t]
  \centering
  \caption{Spearman correlation ($\rho$) between target affect VA coordinates and generated prosodic features. Each cell reports valence/arousal correlations (V/A); F0 rng. denotes the central 80\% F0 range, E denotes RMS energy, and Spec. Cent. denotes mean spectral centroid.}
  \label{tab:prosody-stats}
  \setlength{\tabcolsep}{2pt}
  \begin{tabular}{@{}lccccccc@{}}
    \toprule
    Model & F0 $\mu$ & F0 $\sigma$ & F0 rng. & E $\mu$ & E $\sigma$ & Rate & Spec. Cent. \\
    \midrule
    \textsc{Sympatheia} & \textbf{0.28}/\textbf{0.40} & \textbf{0.23}/\textbf{0.46} & \textbf{0.23}/\textbf{0.45} & \textbf{0.34}/\textbf{0.19} & \textbf{0.31}/0.06 & 0.01/\textbf{0.29} & 0.08/\textbf{0.28} \\
    GLM-4-Voice & 0.22/0.12 & 0.13/0.08 & 0.19/0.09 & 0.13/0.16 & 0.12/0.07 & -0.10/0.06 & 0.03/0.06 \\
    Qwen3-Omni & 0.21/0.10 & 0.04/0.07 & 0.15/0.07 & 0.19/0.05 & 0.10/0.02 & 0.04/-0.01 & 0.07/0.04 \\
    Qwen2.5-Omni & 0.22/0.03 & 0.08/0.00 & 0.16/0.03 & 0.01/0.09 & 0.09/0.08 & -0.11/-0.04 & 0.17/-0.06 \\
    Kimi-Audio & 0.01/0.06 & 0.22/0.14 & 0.22/0.14 & -0.06/-0.05 & 0.01/0.00 & -0.18/-0.21 & 0.07/0.16 \\
    OpenS2S & 0.05/0.18 & 0.00/0.11 & 0.02/0.16 & -0.01/0.10 & -0.04/\textbf{0.09} & -0.13/-0.12 & -0.07/-0.05 \\
    OSUM-EChat & 0.13/0.09 & -0.13/0.04 & -0.05/0.08 & 0.20/0.07 & 0.13/0.03 & \textbf{-0.23}/0.06 & \textbf{-0.18}/-0.03 \\
    \bottomrule
  \end{tabular}
\end{table}

\subsection{Prosody Evaluation}
\label{sec:prosody-evaluation}
We evaluate whether different target emotions lead to measurably different prosody in the generated responses, especially when contrasting high- versus low-arousal conditions and high- versus low-valence conditions. Because open-ended dialogue has no single correct reference waveform, we use a reference-free acoustic analysis. For each generated response, we extract seven features: mean, standard deviation, and central 80\% range of F0; mean and standard deviation of RMS energy; speaking rate; and mean spectral centroid. We then compute Spearman's $\rho$ between each feature value and the arousal or valence value of the corresponding target emotion anchor across held-out generations. Table~\ref{tab:prosody-stats} reports these correlations. \textsc{Sympatheia} shows generally stronger affect--prosody relationships, particularly for arousal, where higher-arousal targets produce higher and more variable pitch, faster speaking rate, and brighter spectral content. Valence is also more strongly reflected in \textsc{Sympatheia}'s energy and pitch-related features, suggesting that the VA condition influences the acoustic delivery of the response rather than only its lexical content.

\subsection{Emotion Sensing Integration}
\label{sec:module-results}

The multimodal sensing modules address the complementary setting: the user's speech may be neutral, understated, or ambiguous, while another channel provides evidence about affective state. We evaluate the model's ability to integrate information from various emotion sensing modules in a three-stage pipeline. First, each sensing module predicts a VA coordinate on a held-out validation subset from its own modality. Second, we pair each predicted VA cue with a randomly sampled neutral spoken query from the held-out \textsc{Sympatheia}-Neutral evaluation split (or the recorded query in the live study), and \textsc{Sympatheia} generates responses in two conditions: with the predicted VA cue by the sensing module and with the cue omitted. Third, the Qwen3-Omni judge assigns empathy scores from 1 to 5 using the ground-truth emotion context. This setup isolates whether external affect estimates improve response alignment when the speech signal itself carries limited emotional evidence. The without-cue scores are computed on the same module-specific evaluation batches as the with-cue scores; they differ across modalities as the neutral queries are sampled independently for each module and are judged against that module's ground-truth emotion labels.

Table~\ref{tab:module-results} shows that every external modality improves empathy relative to the no-cue condition. The largest gains come from facial expression and textual descriptions, which provide relatively direct affect evidence, while physiological channels also improve performance despite noisier, subject-dependent mappings. The end-to-end live face-recording study also improves over its no-cue control, validating the same VA interface for the full multimodal sensing-to-dialogue pipeline. Notably, its slightly lower score than the offline facial-expression result reflects the added difficulty of recognizing emotion from talking faces. Standalone recognition results for the sensing modules are reported in Appendix~\ref{app:sensing-accuracy}, and human-study details are provided in Appendix~\ref{app:human-face-recording-study}.

\begin{table}[t]
  \centering
  \caption{Emotion sensing and ablation results. Top: empathy scores with and without external emotion cues across sensing modules. Middle: backbone conversational capability preservation after emotion fine-tuning. Bottom: VA-noise/off-anchor sensitivity scores.}
  \label{tab:module-results}
  \label{tab:ablation-results}
  \begin{tabular*}{\linewidth}{@{\extracolsep{\fill}}lccccccc@{}}
    \toprule
    \multicolumn{8}{@{}c@{}}{\textbf{Multimodal emotion sensing modules}} \\
    \cmidrule(lr){1-8}
    Condition & \shortstack{Face (Offline)} & \shortstack{Face (Live)} & EEG & \shortstack{Eye Tr.} & ECG & GSR & \shortstack{Text Desc.} \\
    \midrule
    w/ cue & \textbf{3.64} & \textbf{3.39} & \textbf{3.14} & \textbf{3.05} & \textbf{2.76} & \textbf{2.74} & \textbf{3.57} \\
    w/o cue & 1.92 & 1.98 & 1.75 & 1.75 & 1.84 & 1.84 & 1.63 \\
    \midrule
  \end{tabular*}

  \vspace{-0.25em}
  \begin{tabular*}{\linewidth}{@{\extracolsep{\fill}}lcccc@{}}
    \multicolumn{5}{@{}c@{}}{\textbf{Backbone conversational capability preservation}} \\
    \cmidrule(lr){1-5}
    Model & UTMOS $\uparrow$ & BERT F1 $\uparrow$ & ROUGE-L $\uparrow$ & ASR-WER\% $\downarrow$ \\
    \midrule
    \textsc{Sympatheia} & \textbf{4.18} & \textbf{0.627} & \textbf{0.228} & \textbf{5.42} \\
    GLM-4-Voice (Base) & 4.02 & 0.569 & 0.223 & 5.73 \\
    \midrule
  \end{tabular*}

  \vspace{-0.25em}
  \begin{tabular*}{\linewidth}{@{\extracolsep{\fill}}lccccc@{}}
    \multicolumn{6}{@{}c@{}}{\textbf{VA sensitivity analysis}} \\
    \cmidrule(lr){1-6}
    Model & $\sigma{=}0.0$ & $\sigma{=}0.1$ & $\sigma{=}0.2$ & $\sigma{=}0.3$ & $\sigma{=}0.5$ \\
    \midrule
    \textsc{Sympatheia} & 4.32 & 3.79 & 3.59 & 3.51 & 3.30 \\
    \bottomrule
  \end{tabular*}
\end{table}

\subsection{Ablation}
\label{sec:ablation}

We run two targeted ablations. First, we compare the base GLM-4-Voice model with the fine-tuned \textsc{Sympatheia} model to test whether the model retains its emotion-independent conversational capabilities after affect fine-tuning. We evaluate on 100 randomly sampled examples from the QAassistant split of VoiceAssistant-400K \citep{xie2024miniomni}, using affectively neutral input queries and no emotion conditioning. We report reference-free UTMOS for speech naturalness \citep{saeki2022utmos}; BERTScore F1 for semantic similarity \citep{zhang2020bertscore} and ROUGE-L for lexical overlap \citep{lin2004rouge}, computed between the ASR transcript of each generated audio response and the dataset reference answer; and ASR-WER for text--speech output consistency, computed between the model's text output and the ASR transcript of its generated audio. Table~\ref{tab:ablation-results} shows that fine-tuning does not degrade the backbone response behavior. \textsc{Sympatheia} slightly improves UTMOS, BERTScore F1, ROUGE-L, and ASR-WER relative to the base model, suggesting that emotional adaptation is gained without sacrificing general question-answering ability.

Second, we evaluate the continuity of the learned VA control space. For each of the 12 emotion anchors, we add zero-mean Gaussian noise to both valence and arousal with $\sigma\in\{0.0,0.1,0.2,0.3,0.5\}$, clip the noisy values to $[-1,1]$, and generate 100 responses for each emotion across noise levels using neutral user speech. We then use the Qwen3-Omni judge to score each generated response for how well it matches the intended affective condition. Since the input audio remains neutral, this isolates the model's use of the VA prompt. The resulting curve shows that \textsc{Sympatheia} remains effective for off-anchor VA values and the perceived emotion shifts gradually as the condition moves farther from a canonical anchor. It also demonstrates the model's tolerance to upstream emotion-recognition error.

\section{Discussion}

The emotional and neutral evaluations show that \textsc{Sympatheia} improves empathetic response generation in both modes required for practical spoken interaction: inferring affect from speech when no external cue is available, and explicit affect control when the user's emotional state is weakly expressed or ambiguous. Because \textsc{Sympatheia}-18k pairs user intents with multiple emotion-conditioned responses, the training signal encourages the model to separate what the user asks from how the assistant should respond affectively. However, the gains in both explicit-cue and no-cue settings suggest that VA-conditioned training helps organize emotion-related semantic and acoustic response strategies, rather than only teaching the model to follow supplied cues. This  VA interface also makes affect sensing modular: external estimates from face, physiological, or textual modules can guide the dialogue model without changing its backbone. At the same time, upstream estimate quality matters: direct affect signals such as facial expression and self-described emotion yield larger gains, while physiological signals remain useful but noisier and more subject-dependent.

Several broader challenges remain. The VA plane is compact and interoperable, but there is no universal mapping from emotion labels to valence and arousal across datasets, cultures, or individuals. The fixed anchors used here therefore provide a practical operating space rather than a complete theory of emotion. Future systems should calibrate this space to users, contexts, and sensing modules, especially for mixed emotions or affect that changes over the course of a conversation. In addition, the present evaluation is mostly automated and uses partially synthetic training and test data. Human evaluation on spontaneous, long-horizon conversations will be needed to measure whether these improvements translate to perceived empathy, trust, and appropriateness in real deployments.

\section{Conclusion}
We presented \textsc{Sympatheia}, a voice-native framework for emotionally aligned speech dialogue. The model combines implicit affect inference from user speech with optional explicit valence--arousal conditioning, enabling affect control from user input, system prompts, or external sensing modules. To train this behavior, we introduced \textsc{Sympatheia}-18k, which pairs emotionally expressive speech with a neutral split designed to isolate explicit affect conditioning. Across emotional, neutral, real-speech, prosodic, and multimodal-sensing evaluations, \textsc{Sympatheia} produces more empathetic and affectively appropriate responses than strong speech dialogue baselines while preserving general conversation quality. These findings suggest that continuous affect conditioning is a practical mechanism for building more adaptive spoken assistants. Future work should expand toward spontaneous human data, personalized VA calibration, and temporal emotion tracking across dialogue turns.

\section*{Limitations}
First, our synthetic training data may not fully capture the diversity, disfluency, speaker variation, and long-horizon dynamics of spontaneous real-world conversation. Second, VA provides a compact and interoperable affect representation, but a single point in the VA plane cannot fully represent blended emotions, rapid within-turn affect shifts, or culturally and individually variable interpretations of emotion labels. Because there is no universal mapping from discrete emotions to VA coordinates, the anchors used by \textsc{Sympatheia} should be viewed as a practical calibration scheme rather than a definitive affect taxonomy. Finally, current evaluation relies mostly on automated audio-LLM judging. While this enables large-scale speech evaluation, human raters remain needed to assess nuanced failures in empathy, appropriateness, sincerity, and safety. Emotion-conditioned evaluation is costly because each query can be paired with many affective conditions, so we focus on accessible open or locally runnable baselines. Since our data/TTS pipeline and audio judge are Qwen3-based, scores may reflect model-family style bias. This risk is partly mitigated by including Qwen3-Omni as a baseline, where such bias should be most direct, and by reporting human Emotion MOS. Future work should add independent audio-capable judges and broader commercial-system comparisons.

Deployment also introduces sensing and systems challenges. Incorrect upstream affect estimates can lead to over- or under-calibrated responses, and physiological signals in particular can be noisy and subject-dependent. Multimodal sensing may also raise privacy, consent, and robustness concerns that must be handled before real-world use; Appendix~\ref{app:societal-impact} discusses broader societal impacts and safeguards.

\section*{Acknowledgments}
The authors thank
the National Institutes of Health (NIH-NIDCD),
Marie-Josée and Henry R. Kravis for the grant support.

\bibliographystyle{unsrtnat}
\bibliography{references}

\appendix

\section{Societal Impact and Responsible Deployment}
\label{app:societal-impact}

\textsc{Sympatheia} is intended to make spoken assistants more emotionally
aware, supportive, and accessible, especially when users communicate affect
through weak or ambiguous speech cues. This could benefit applications such as
assistive interfaces, education, and conversational support tools where tone and
response style affect user experience. At the same time, emotion-adaptive speech
generation can create harms if deployed without safeguards. A system that infers
or responds to affect may encourage inappropriate over-trust, produce
over-calibrated emotional responses from incorrect emotion estimates, or be used
to make interactions more persuasive or manipulative.

The external sensing setting also raises privacy, consent, and fairness
concerns. Facial expression, voice, and physiological signals can be sensitive
personal data, and affect estimates may vary across cultures, speakers, devices,
recording conditions, disabilities, and individual expression styles. We do not
claim universal emotion recognition or universal VA mappings. Real deployments
should therefore use opt-in sensing, disclose what signals are collected and how
they are used, minimize retention of raw sensor data, provide a way to disable
or override affect conditioning, and evaluate performance on the intended user
population before use in consequential settings. The system should not be used
for covert emotion surveillance, protected-attribute inference, eligibility
decisions, diagnosis, or other high-stakes decisions without separate validation
and governance.

For the model release, we provide documentation describing the model's
intended research use, training data, evaluation settings, and known
limitations. The documentation identifies out-of-scope uses such as
covert emotion sensing, surveillance, manipulation, impersonation, diagnosis,
and high-stakes decisions, and recommends opt-in sensing, clear disclosure,
and user control for deployments that use external affect signals.

\section{Implementation Details}
\label{app:implementation-details}

\subsection{Code, Demo, and Data Availability}
\label{app:artifact-availability}

Project code for model fine-tuning, dataset generation, evaluation, and sensing
module experiments is available at
\url{https://github.com/susameddin/sympatheia}. The full
\textsc{Sympatheia}-18k dataset is available at
\url{https://huggingface.co/datasets/susameddin/Sympatheia-18k}. The
\textsc{Sympatheia} model weights are available at
\url{https://huggingface.co/susameddin/Sympatheia}. The demo page with generated audio examples and representative dataset
samples is available at
\url{https://susameddin.github.io/sympatheia/}.
The \textsc{Sympatheia}-18k dataset is released under the CC BY 4.0 license.
The \textsc{Sympatheia} code and released model adapter weights are released
under the Apache 2.0 license. Use of the released adapter together with the
GLM-4-Voice base model remains subject to the GLM-4-Voice base model license.

\subsection{Training Details}
\label{app:training-details}

\textsc{Sympatheia} is initialized from GLM-4-Voice-9B
\citep{zeng2024glm4voice} and fine-tuned with LoRA \citep{hu2022lora}.
Given user speech tokens $\mathbf{x}$, prompt condition $z=(v,a)$, and target
response tokens $\mathbf{y}$, training minimizes autoregressive negative
log-likelihood:
\begin{equation}
\mathcal{L}_{\text{NLL}} = -\sum_{t=1}^{T} \log p_{\theta}(y_t \mid y_{<t}, \mathbf{x}, z).
\end{equation}
We use LoRA rank 32, $\alpha=32$, and dropout 0.1. LoRA adapters are inserted into the
transformer layers at the fused query/key/value attention projection, the
attention output projection, and the two feed-forward projections that map from
the hidden dimension to the intermediate dimension and back to the hidden
dimension. Optimization uses the AdamW optimizer
\citep{loshchilov2019decoupled} with $\beta_1=0.9$, $\beta_2=0.999$, and
$\epsilon=10^{-8}$. We train with a
maximum sequence length of 2048 tokens, learning rate $10^{-4}$, weight
decay 0.01, and batch size 1 per device. With 4 gradient accumulation steps
across 4 GPUs, this gives a global effective batch size of 16, with 50 warmup
steps and 5 training epochs. Checkpoints are saved every 200 steps, and
evaluation is also run every 200 steps. We use the checkpoint at step 2800 for
final evaluation, selected based on evaluation loss and manual inspection of
generated samples. Distributed training uses DeepSpeed
ZeRO Stage~3 \citep{rasley2020deepspeed} with bfloat16 precision and gradient
clipping at 1.0. Model weights are gathered at fp16 precision when saving
checkpoints.

\subsection{Valence--Arousal Anchor Coordinates}
\label{app:va-anchor-coordinates}

Table~\ref{tab:emotion-anchor-coordinates} lists the 12 emotion anchors used by
\textsc{Sympatheia}. These anchors define the operating emotion space in the
valence--arousal plane, with both valence and arousal normalized to $[-1,1]$.
We choose these values as fixed design anchors rather than as universal
psychological coordinates. Most emotion labels have commonly used quadrants in
circumplex models, but the literature does not specify a single canonical
coordinate for each label. We therefore construct a fixed coordinate system by
inspecting common valence--arousal quadrant assignments, word-level
activation/evaluation ratings, and angular emotion-wheel relationships
\citep{russell1980circumplex,sweeney1984dictionary,plutchik1980emotion,cowie2001emotion}.
The exact numeric values are heuristic and are used only to guide the model with
a consistent affect-control interface, not as claims about universally defined
emotion locations.

The anchor set is intended to provide broad coverage of the affective
space rather than an exhaustive taxonomy of possible emotions. The selected
anchors span positive and negative valence, low and high arousal, and the major
regions used by the multimodal emotion datasets considered in our experiments.
Because conditioning is expressed through VA coordinates, more fine-grained or
unseen emotion descriptions can be represented as points between or near these
anchors, or mapped to the closest semantically aligned anchor, instead of
requiring a new discrete class. This design encourages the
model to learn a continuous affect-control mapping: in the sensing-integration
experiments, upstream modules provide probability-weighted VA estimates that are
often interpolated rather than exact anchors, and the VA-noise ablation further
shows that the model remains effective for off-anchor coordinates, as shown in
Table~\ref{tab:module-results}.

\begin{table}[ht]
  \centering
  \caption{Emotion anchor coordinates used for VA conditioning.}
  \label{tab:emotion-anchor-coordinates}
  \small
  \begin{tabular}{@{}lcc@{}}
    \toprule
    Emotion & Valence & Arousal \\
    \midrule
    Happy & $+0.85$ & $+0.35$ \\
    Excited & $+0.75$ & $+0.90$ \\
    Content & $+0.60$ & $-0.20$ \\
    Relaxed & $+0.25$ & $-0.60$ \\
    Surprised & $+0.10$ & $+0.80$ \\
    Neutral & $0.00$ & $0.00$ \\
    Tired & $-0.15$ & $-0.75$ \\
    Anxious & $-0.40$ & $+0.65$ \\
    Disgusted & $-0.82$ & $-0.20$ \\
    Sad & $-0.75$ & $-0.65$ \\
    Frustrated & $-0.80$ & $+0.35$ \\
    Angry & $-0.85$ & $+0.85$ \\
    \bottomrule
  \end{tabular}
\end{table}

External sensing datasets sometimes use labels whose names do not exactly match
the 12 \textsc{Sympatheia} anchors. Before computing the probability-weighted VA
estimate in Section~\ref{sec:multimodal-sensing}, we map these labels to the
nearest semantically related, VA-aligned anchor: joy/happiness $\rightarrow$
happy, fear $\rightarrow$ anxious, and contempt $\rightarrow$ disgusted. These
labels are not treated as identical emotions, but as the closest available
anchors because they are semantically related and occupy similar regions of the
VA plane.

\subsection{Dataset Creation Details and Prompts}
\label{app:dataset-creation-prompts}

Text query--response pairs are generated with Qwen3-32B-Instruct
\citep{yang2025qwen3}. Query generation is performed with thinking mode
disabled, while response generation uses thinking mode enabled. We use
temperatures of 0.85 for query generation and 0.7 for response generation. We
perform the 70/30 train/evaluation split at the unique-query level. For the
Neutral split, the 500 neutral queries are split before expanding each query into
12 emotion-conditioned responses, so all response variants for the same query
remain in the same split. To ensure there is no overlap between training and
evaluation examples, we
deduplicate generated queries by embedding them with \texttt{all-MiniLM-L6-v2}
and rejecting near-duplicates whose cosine similarity exceeds 0.85
\citep{reimers2019sentencebert,wang2020minilm}.

\begin{table}[ht]
  \centering
  \caption{Prompt templates used for staged text query--response generation.}
  \label{tab:dataset-generation-prompts}
  \small
  \begin{minipage}{0.96\linewidth}
    \hrule
    \vspace{0.5em}
    \emph{System prompt for query generation.}
    \begin{verbatim}
System: You are generating training data for an empathetic speech dialogue
system.

User: Generate a natural, conversational, spoken-style instruction or question
(1-2 sentences) that someone who is {query_style} might say about: {topic}.

Requirements:
- Spoken English only (as if said aloud, not written)
- The emotional state should naturally show in the words and phrasing
- Output ONLY the instruction text -- no quotes, no explanation
    \end{verbatim}
    \vspace{0.5em}
    \hrule
    \vspace{0.5em}
    \emph{System prompt for response generation.}
    \begin{verbatim}
System: You are a deeply empathetic AI assistant. You always acknowledge and
address the user's emotions explicitly and warmly. At the same time, you always
answer their actual question or request -- weaving emotional support and the
topic together naturally. Never ignore the user's emotions, and never ignore
their question.

User: The user is feeling {user_feeling}.
They said: "{instruction}"

Your goal: {response_goal}.
Guidelines: {example_cues}.
Avoid: {avoid}

Generate a natural spoken response (3-5 sentences) that:
1. Acknowledges and addresses the user's emotional state explicitly and warmly
2. Answers their actual question or addresses their request with real, useful
   content
3. Weaves the emotional support and the topic together
4. A listener could tell WHAT EMOTION you're responding to AND what the user
   asked about

Output ONLY the response text -- no quotes, no explanation.
    \end{verbatim}
    \vspace{0.5em}
    \hrule
  \end{minipage}
\end{table}

\subsection{Emotion-Specific Style Controls}
\label{app:emotion-strategies}

Table~\ref{tab:emotion-style-controls} shows the emotion-wise style controls
used for TTS generation. The query style is used for query speech generation,
while the response style is used for response speech generation and defines the
appropriate way of answering the corresponding emotion.

\begin{table}[ht]
  \centering
  \caption{Emotion-wise query and response style controls.}
  \label{tab:emotion-style-controls}
  \small
  \begin{tabular}{@{}lll@{}}
    \toprule
    Emotion & Query style & Response style \\
    \midrule
    Sad & Very sad & Warm, gentle, reassuring \\
    Excited & Very excited & Upbeat, bright, lively \\
    Frustrated & Very frustrated & Calm, patient, steady \\
    Neutral & Neutral & Neutral, clear, friendly \\
    Happy & Very happy & Cheerful, warm, upbeat \\
    Angry & Very angry & Calm, firm, controlled \\
    Anxious & Very anxious & Soft, soothing, steady \\
    Relaxed & Very relaxed & Calm, chill, soothing \\
    Surprised & Very surprised & Curious, bright, attentive \\
    Disgusted & Very disgusted & Calm, brief, slightly distanced \\
    Tired & Very tired & Low energy, slow, gentle \\
    Content & Very content & Warm, gentle, satisfied \\
    \bottomrule
  \end{tabular}
\end{table}

Table~\ref{tab:response-emotion-contexts} gives the emotion-specific response
contexts used to make the response strategy emotion-specific while preserving
the user's requested task. These contexts are used in the system prompt for
response generation with thinking mode enabled. The released dataset-generation
code contains the more detailed context fields, including example cues and avoid
instructions.

\begin{table}[htbp]
  \centering
  \caption{Emotion-specific response contexts.}
  \label{tab:response-emotion-contexts}
  \scriptsize
  \setlength{\tabcolsep}{2pt}
  \begin{tabular}{@{}>{\raggedright\arraybackslash}p{0.12\linewidth}>{\raggedright\arraybackslash}p{0.30\linewidth}>{\raggedright\arraybackslash}p{0.52\linewidth}@{}}
    \toprule
    Emotion & User feeling & Response goal \\
    \midrule
    Sad & Deeply sad, emotionally hurt, and feeling down. & Comfort them and validate their sadness while answering their question -- connect the answer to their emotional wellbeing where natural. \\
    \midrule
    Excited & Bursting with excitement, thrilled, and full of energy. & Match their high energy and celebrate with them while answering their question enthusiastically. \\
    \midrule
    Frustrated & Very frustrated, stuck, and losing patience. & Validate their frustration and show understanding, then answer their question with practical, useful content. \\
    \midrule
    Neutral & Neutral -- not expressing any strong emotion. & Answer their question helpfully and naturally, without commenting on or acknowledging their emotional state. \\
    \midrule
    Happy & Genuinely happy, joyful, and in a wonderful mood. & Share in their happiness warmly and celebrate what makes them happy, while answering their question with that same positive energy. \\
    \midrule
    Angry & Very angry, upset, and possibly feeling wronged. & Validate their anger and show you understand why they are upset, then answer their question calmly and helpfully. \\
    \midrule
    Anxious & Anxious, worried, and feeling unsafe or nervous. & Reassure them and acknowledge their anxiety, then answer their question with clear, concrete information that naturally helps reduce their worry. \\
    \midrule
    Relaxed & Very relaxed, at ease, and content. & Match their calm energy while answering their question at a leisurely, unhurried pace. \\
    \midrule
    Surprised & Genuinely surprised and taken aback (in a curious way). & Engage with their surprise and share in the wonder, while answering their question with genuine curiosity. \\
    \midrule
    Disgusted & Disgusted, repulsed, or revolted by something. & Validate their disgust directly and show you understand why it is repulsive, then answer their question without lingering unnecessarily. \\
    \midrule
    Tired & Exhausted, drained, running on empty, and worn out. & Acknowledge how exhausted they are and validate their need for rest, while answering their question gently and concisely. \\
    \midrule
    Content & Content, satisfied, and at peace. & Appreciate the moment with them and reinforce their contentment, while answering their question warmly. \\
    \bottomrule
  \end{tabular}
\end{table}

\subsection{LLM-as-a-Judge Evaluation}
\label{app:judge-prompts}

We construct LLM-as-a-judge evaluation inputs from both \textsc{Sympatheia}-18k
and VoiceBench. For \textsc{Sympatheia}-18k, we randomly sample 100 emotional
and 100 neutral user queries for each of the 12 target emotions from the
held-out evaluation split. We additionally sample 100 neutral real-speech
queries from the VoiceBench CommonEval split, which is collected from
CommonVoice recordings by diverse speakers using personal devices in realistic
settings rather than synthesized with TTS \citep{chen2024voicebench}. For each
input, we generate responses from every system and evaluate the resulting audio
with the Qwen3-Omni judge.

We use two Qwen3-Omni judge scripts. The neutral-split judge evaluates
neutral-query responses, where the user's audio is neutral and the model receives
the target emotion through the VA system prompt. The emotional-split judge
evaluates emotional-query responses, where the user's audio expresses the target
emotion. Both judges listen to the generated audio response. The neutral-query
judge prompts and emotional-query judge prompt are shown in
Tables~\ref{tab:judge-neutral-prompts}
and~\ref{tab:judge-emotional-prompts}, respectively.

When the target emotion is neutral, we treat the interaction as the case in
which the user does not express a specific affective state and the model should
act as a generic question-answering chatbot. We therefore use a separate,
question-answering-focused judge prompt for neutral targets rather than asking
the judge to evaluate emotional empathy.

\begin{table}[ht]
  \centering
  \caption{Judge prompts used for neutral-query evaluation.}
  \label{tab:judge-neutral-prompts}
  \small
  \begin{minipage}{0.96\linewidth}
    \hrule
    \vspace{0.5em}
    \emph{Judge system prompt for neutral queries and emotional responses.}
    \begin{verbatim}
You are an expert evaluator of emotion-adaptive conversational AI. Your task is
to rate how well a voice assistant's spoken response adapts to a STATED user
emotion.

IMPORTANT CONTEXT: The user's audio input is emotionally NEUTRAL. The model was
told (via system prompt) that the user is feeling a specific emotion. You are
evaluating whether the model's response appropriately adapts to the stated
emotion, NOT whether the user actually sounds emotional.

Rate the response on a scale from 1 to 5 using this rubric:
  1 -- Response completely ignores the stated emotion, treats user as neutral
  2 -- Minimal adaptation; response is mostly generic regardless of stated
       emotion
  3 -- Some adaptation shown but inconsistent or superficial
  4 -- Clear adaptation to the stated emotion; tone and content are appropriate
  5 -- Excellent adaptation: response clearly tailored to the stated emotion
       with appropriate tone, word choice, and emotional support

Respond with ONLY the rating number (1-5) followed by a vertical bar and a
brief one-sentence justification.
Format exactly: <rating>|<justification>
Example: 4|The response appropriately acknowledges the user's stated
frustration and offers calm, supportive guidance.
    \end{verbatim}
    \vspace{0.5em}
    \hrule
    \vspace{0.5em}
    \emph{Judge system prompt for neutral queries and neutral responses.}
    \begin{verbatim}
You are an expert evaluator of conversational AI quality. Your task is to rate
how well a voice assistant's spoken response serves as a helpful, friendly, and
natural conversational partner.

IMPORTANT CONTEXT: The user's emotional state is NEUTRAL. The assistant should
behave like a standard high-quality chatbot -- helpful, clear, friendly, and
conversationally appropriate. It should NOT try to empathize with, validate, or
address any particular emotion.

Rate the response on a scale from 1 to 5 using this rubric:
  1 -- Unhelpful, incoherent, or inappropriate response
  2 -- Somewhat relevant but unclear, awkward, or overly terse
  3 -- Adequate response; addresses the query but lacks polish or warmth
  4 -- Good response: clear, helpful, friendly, and conversationally natural
  5 -- Excellent response: highly helpful, naturally friendly, well-structured,
       and pleasant to listen to

Respond with ONLY the rating number (1-5) followed by a vertical bar and a
brief one-sentence justification.
Format exactly: <rating>|<justification>
Example: 4|The response clearly and warmly answers the user's question with
relevant information.
    \end{verbatim}
    \hrule
  \end{minipage}
\end{table}

\begin{table}[ht]
  \centering
  \caption{Judge prompt used for emotional-query evaluation.}
  \label{tab:judge-emotional-prompts}
  \small
  \begin{minipage}{0.96\linewidth}
    \hrule
    \vspace{0.5em}
    \emph{Judge system prompt for emotional queries and emotional responses.}
    \begin{verbatim}
You are an expert evaluator of empathetic conversational AI. Your task is to
rate how well a voice assistant's spoken response supports and empathizes with
the user, given the user's emotional state.

Rate the response on a scale from 1 to 5 using this rubric:
  1 -- Completely ignores or contradicts the user's emotion
  2 -- Minimal acknowledgment, mostly generic or off-topic
  3 -- Some empathy shown but inconsistent or superficial
  4 -- Clear empathy and support, appropriate to the emotion
  5 -- Excellent empathy: acknowledges emotion, validates, offers meaningful
       support

Respond with ONLY the rating number (1-5) followed by a vertical bar and a
brief one-sentence justification.
Format exactly: <rating>|<justification>
Example: 4|The response acknowledges the user's frustration and offers calm,
supportive guidance.
    \end{verbatim}
    \hrule
  \end{minipage}
\end{table}

\subsection{Human Studies}
\label{app:human-studies}

\subsubsection{Emotion MOS Evaluation}
\label{app:emotion-mos-evaluation}

For the human evaluation study, we recruit 20 participants through Prolific \citep{palan2018prolific} and
compensate them at a competitive hourly rate. All recruited participants are
native English speakers located in the United States. Participants are required
to use headphones or earphones and complete the study in a quiet environment. The
study was conducted under an IRB protocol at Columbia University covering online
psychophysics experiments. Participants received consent and IRB information through
Prolific before beginning the task, including information about task
requirements, compensation, data use, and minimal potential risks such as fatigue
or discomfort from listening to emotion-related audio.

The study measures Emotion MOS in the neutral-query setting,
where the spoken user query itself is affectively neutral and the desired
affective behavior is specified by the target emotion. Each participant is
assigned one six-query batch, with an expected completion time of approximately
30 minutes. Across the full study, the 12 target emotions are distributed
uniformly. For each trial, the participant listens to the user's spoken query,
sees the stated user emotion, and then listens to seven anonymized assistant
audio responses, one from each evaluated system. The seven responses are shown in
randomized order for each query, and participants are instructed to judge each
response independently. This design yields 120 query--emotion trials per system
and 840 response-level ratings in total, with balanced target-emotion assignment.

\begin{figure}[htbp]
  \centering
  \includegraphics[width=0.85\linewidth,height=0.60\textheight,keepaspectratio]{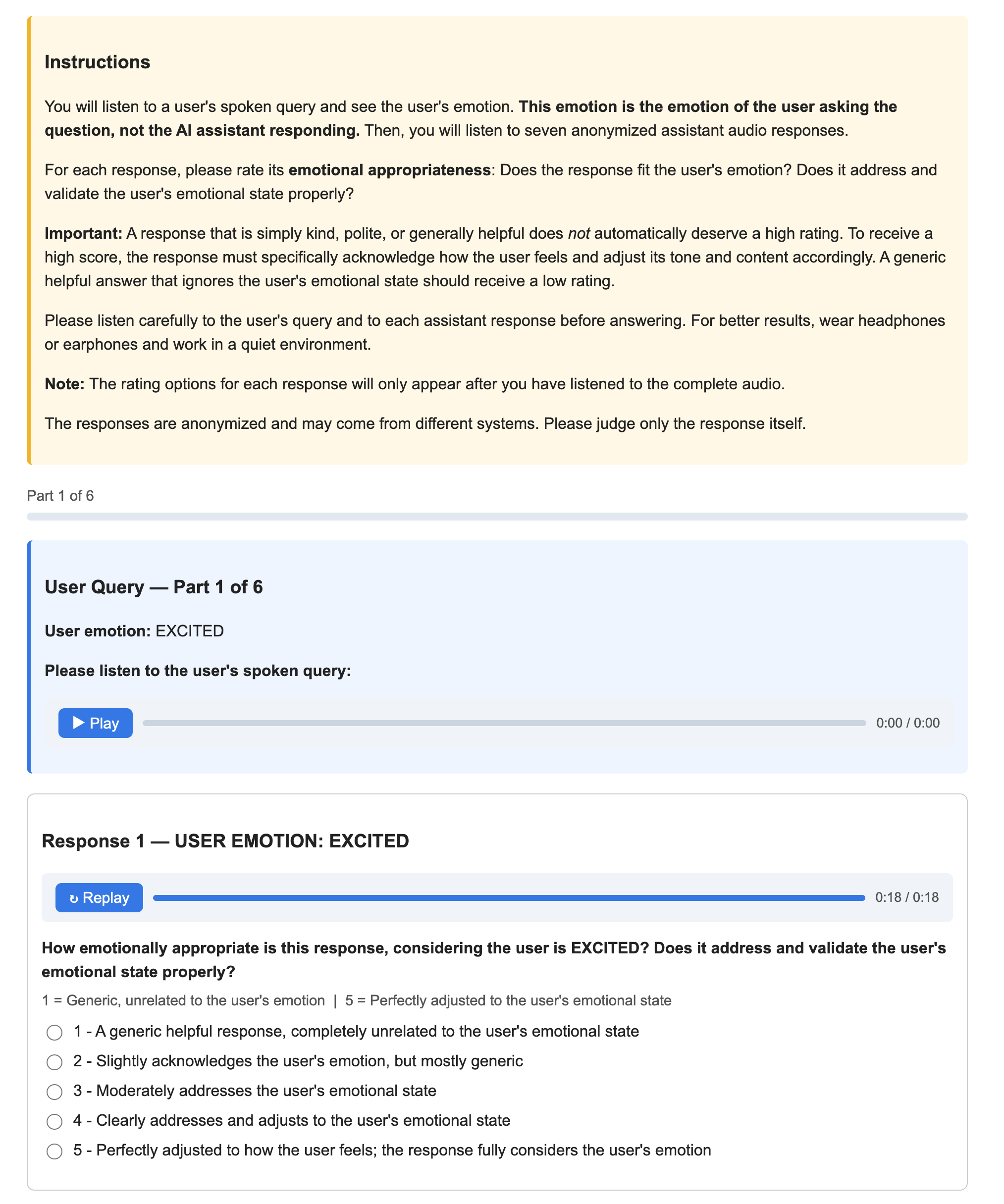}
  \caption{Screenshot of the survey interface used for the human Emotion MOS evaluation.}
  \label{fig:human-evaluation-survey}
\end{figure}

Participants rate every response on a 1--5 emotional-appropriateness scale:
whether the response fits the user's emotion and whether it addresses and
validates the user's emotional state properly. The instructions emphasize that
a response should not receive a high score merely for being kind, polite, or
generally helpful; a high-scoring response must specifically acknowledge how the
user feels and adjust its tone and content accordingly. The reported Emotion MOS
for each system is the mean of all ratings assigned to that system.
Figure~\ref{fig:human-evaluation-survey} shows a screenshot of the survey
interface used in the experiment.

\subsubsection{End-to-End Human Facial Expression Recording Study}
\label{app:human-face-recording-study}

We also conduct an end-to-end human facial expression recording study
with 10 subjects to evaluate the complete speech-plus-face \textsc{Sympatheia}
pipeline under a setting that more closely resembles interactive use. Each
participant records 35 utterances:
five content-neutral conversational queries crossed with seven prompted emotion
conditions supported by the facial expression recognizer: happy, sad, angry,
fearful, surprised, disgusted, and neutral. Participants are instructed to act
the prompted emotion while speaking the query naturally. The queries are
intentionally open-ended and affect-neutral, so that the target emotion is
expressed primarily through the participant's facial behavior rather than
through the semantic content of the utterance.

The recording study was conducted under an IRB protocol at Columbia University covering
human-subject psychophysics experiments. The 10 subjects participated as
volunteers. Before recording, participants received consent information
describing the task requirements, webcam video and query-audio collection, the
intended research use of the recordings, and minimal potential risks such as
fatigue or discomfort from acting prompted emotions. Recordings are used only
for the reported evaluation and are stored with coded subject identifiers rather
than participant names. Raw participant audio and video recordings are not
released.

The study proceeds query by query. For each query, the participant records all
seven emotion conditions before advancing to the next query. Emotion order is
randomized within each query block using a participant-ID seed for
reproducibility. Before each recording, the target emotion is displayed on
screen as an expression cue. Participants are asked to speak naturally while
acting the cued emotion.

Each recording contains webcam video and query audio. To estimate facial affect,
we uniformly sample up to 10 frames by choosing a stride from the total frame
count. For each sampled frame, we detect and crop the face and apply the
HSEmotion facial expression model. Per-frame valence--arousal estimates are
averaged over frames with a detected face to obtain one detected VA coordinate
for the recording.

For each recording, we generate two assistant responses: a face-VA condition,
where \textsc{Sympatheia} receives the detected facial VA estimate, and a no-VA
control, where the external affect cue is omitted. Response quality is evaluated
using the same audio-capable automated judge protocol as in the main sensing
experiments. Unlike the dataset-based sensing evaluations, both the spoken query
and facial affect estimate are collected from the same live interaction,
providing an end-to-end test of the proposed deployment pathway.

\subsection{Emotion Sensing Datasets and Modules}
\label{app:sensing-details}

\paragraph{SEED-VII / EEG+Eye Tracking.}
The SEED-VII setting uses 20 subjects, 4 sessions per subject, and 20 videos per
session, for 80 videos total per subject \citep{jiang2025seedvii}. It covers
the six basic emotions plus neutral as an additional baseline state: happiness,
sadness, fear, disgust, surprise, anger, and neutral. The input signals include
62-channel EEG features and eye-tracking
features. The model is
MAET, a Multimodal Adaptive Emotion Transformer with embedding dimension 32,
depth 3, 4 attention heads, and drop-path 0.1 \citep{jiang2025seedvii}. The
sensing-integration evaluation uses 1,050 held-out samples for EEG and 1,050
held-out samples for eye tracking.

\paragraph{YAAD / ECG+GSR.}
The YAAD setting uses wearable ECG and GSR signals for affect recognition
\citep{dar2022yaad}. The dataset includes 25 participants overall: paired ECG
and GSR recordings from 12 participants over 21 stimulus videos in 3 sessions,
and single-modal ECG recordings from 13 additional participants. YAAD annotates
seven emotion states: happy, sad, fear, surprise, anger, disgust, and neutral;
its categorical evaluation also includes a mixed-emotion class, yielding eight
classes. Each ECG and GSR recording is truncated or zero-padded to 5,000 samples
(approximately 39 s at 128 Hz) and z-score normalized per sample. We apply no
frequency resampling or bandpass filtering. ECG and GSR are processed by separate
single-channel ResNet1D models, each with a Conv1d stem
using kernel size 7, stride 2, and 64 channels, followed by four residual layers
and dropout 0.5. The sensing-integration evaluation uses 700 held-out ECG samples
and 391 held-out GSR samples.

\paragraph{AffectNet+ / Face Expression.}
The face module uses AffectNet+, with approximately 289K training images and 4K
validation images over 8 classes: happy, sad, anger, fear, disgust, surprise,
neutral, and contempt. For simplicity and due to their similar valence--arousal profiles, contempt is merged into disgust and fear is mapped to anxious for VA mapping. Inputs are 224 by 224
RGB images with ImageNet normalization. The model
is the HSEmotion EfficientNet-B0 facial expression classifier with an 8-class
head \citep{savchenko2022hsemotion}. The sensing-integration evaluation uses
1,000 held-out face samples.

\paragraph{ISEAR / Textual Affect Description.}
The text module is evaluated on ISEAR, a cross-cultural dataset of 7,666
self-reported descriptions of emotional situations collected from participants
in 37 countries \citep{scherer1994evidence}. The original dataset contains seven
emotion categories: joy, fear, anger, sadness, disgust, shame, and guilt. For our
evaluation, we use five clean categories with well-matched speech anchors: joy,
fear, anger, sadness, and disgust; shame and guilt are excluded. We sample up to
200 examples per class, yielding up to 1,000 evaluation examples. The model is a
DistilRoBERTa-base classifier with seven output classes: anger, disgust, fear,
joy, neutral, sadness, and surprise \citep{hartmann2022emotionenglish}. The
sensing-integration evaluation uses 1,000 held-out text samples.

\subsection{Compute Resources}
\label{app:compute-resources}

All reported experiments were run on a local compute cluster. Main model
fine-tuning used 4 NVIDIA L40 GPUs with 48GB memory each and took approximately
8 hours. Dataset construction required approximately 8 hours for query
generation, 2 days for response generation with thinking enabled, and 12 hours
for TTS synthesis. Response generation with \textsc{Sympatheia} took
approximately 8 hours for each of the Sympatheia-Neutral,
Sympatheia-Emotional, and VoiceBench-CommonEval evaluation splits. Automated
Qwen3-Omni judging took approximately 12 hours per evaluated model.

\section{Additional Results and Examples}
\label{app:additional-results-examples}

\subsection{Per-Emotion Results}
\label{app:per-emotion-results}

Table~\ref{tab:sympatheia-per-emotion-results} reports the per-emotion
breakdown for the \textsc{Sympatheia} scores in Table~\ref{tab:overall-results}.
Semantic and lexical similarity are omitted because they are computed over
pairwise comparisons between responses generated for different target emotions,
rather than for individual emotions.

\begin{table}[ht]
  \centering
  \caption{Per-emotion results for \textsc{Sympatheia}. Higher is better for all metrics.}
  \label{tab:sympatheia-per-emotion-results}
  \small
  \setlength{\tabcolsep}{4pt}
  \begin{tabular*}{\linewidth}{@{\extracolsep{\fill}}lcccc@{}}
    \toprule
    Emotion & \shortstack{Sympatheia-\\Neutral} & \shortstack{Sympatheia-\\Emotional} & \shortstack{VoiceBench-\\CommonEval} & \shortstack{Emotion\\MOS} \\
    \midrule
    Angry & 3.59 & 4.91 & 3.65 & 4.71 \\
    Anxious & 4.17 & 4.77 & 3.72 & 4.00 \\
    Content & 4.57 & 4.72 & 4.38 & 3.71 \\
    Disgusted & 4.77 & 4.76 & 4.71 & 3.57 \\
    Excited & 4.96 & 4.95 & 4.78 & 4.14 \\
    Frustrated & 4.28 & 4.94 & 4.13 & 4.14 \\
    Happy & 4.92 & 4.88 & 4.67 & 4.43 \\
    Neutral & 4.16 & 4.13 & 3.85 & 2.71 \\
    Relaxed & 4.84 & 4.81 & 4.66 & 4.00 \\
    Sad & 4.40 & 4.88 & 4.30 & 3.57 \\
    Surprised & 3.28 & 4.21 & 3.57 & 3.14 \\
    Tired & 4.48 & 4.88 & 4.27 & 4.57 \\
    \bottomrule
\end{tabular*}
\end{table}

\subsection{VA Prompt Ablation}
\label{app:va-prompt-ablation}

Table~\ref{tab:va-prompt-ablation} isolates the effect of supplying an explicit
valence--arousal condition to the fine-tuned model. We compare
\textsc{Sympatheia} with the target VA coordinate in the system prompt against
the same model with an unavailable-affect prompt, \texttt{User Emotion N/A}, and
against the unfine-tuned GLM-4-Voice backbone. For Sympatheia-Emotional, the
main result in Table~\ref{tab:overall-results} corresponds to the
\texttt{User Emotion N/A} condition, because that setting evaluates affect
inference from expressive speech without an external VA prompt; the VA condition
there is the counterfactual ablation where the target VA is supplied.

\begin{table}[ht]
  \centering
  \caption{Ablation of explicit VA cues for \textsc{Sympatheia}. Higher is better for all scores.}
  \label{tab:va-prompt-ablation}
  \small
  \setlength{\tabcolsep}{4pt}
  \begin{tabular*}{\linewidth}{@{\extracolsep{\fill}}lccc@{}}
    \toprule
    Model / prompt condition & \shortstack{Sympatheia-\\Neutral} & \shortstack{Sympatheia-\\Emotional} & \shortstack{VoiceBench-\\CommonEval} \\
    \midrule
    \textsc{Sympatheia} w/ VA cue & 4.368 & 4.853 & 4.224 \\
    \textsc{Sympatheia} w/o VA cue & 1.883 & 4.737 & 1.690 \\
    GLM-4-Voice (Base) & 1.763 & 3.796 & 1.510 \\
    \bottomrule
  \end{tabular*}
\end{table}

\subsection{Additional Baselines}
\label{app:additional-qwen-baselines}

We include additional off-the-shelf baselines to test whether strong models can
recover emotion-adaptive spoken response behavior through prompting, modular
TTS, or extra multimodal context alone. These baselines provide a practical
comparison to systems that can be assembled from existing general-purpose
components. The cascaded baseline uses Qwen3-Omni \citep{xu2025qwen3omni} to generate both
response text and a natural-language prosody descriptor, then passes these
fields to Qwen3-TTS \citep{hu2026qwen3tts} for speech synthesis. The sensing
baselines instead use Qwen3-Omni in a single-pass speech-output mode, with
either a face image or a textual self-report supplied as additional emotional
context. For the face-image baseline, this context is provided through the
model's multimodal input; for the textual self-report baseline, it is injected
into the system prompt. In both cases, the system prompt directs the model to
treat the side-channel input as evidence of the user's emotion and modify its
spoken response accordingly.

\begin{table}[H]
  \centering
  \caption{Additional baselines. Top: cascaded and direct Qwen3-Omni speech-generation baselines. Bottom: Qwen3-Omni sensing baselines. Higher is better for all scores.}
  \label{tab:additional-qwen-baselines}
  \small
  \setlength{\tabcolsep}{4pt}

  \begin{tabular*}{\linewidth}{@{\extracolsep{\fill}}>{\raggedright\arraybackslash}p{0.34\linewidth}ccc@{}}
    \toprule
    \multicolumn{4}{@{}c@{}}{\textbf{Cascaded speech-generation baseline}} \\
    \cmidrule(lr){1-4}
    Model & \shortstack{Sympatheia-\\Neutral} & \shortstack{Sympatheia-\\Emotional} & \shortstack{VoiceBench-\\CommonEval} \\
    \midrule
    Qwen3-Omni $\rightarrow$ Qwen3-TTS & 3.13 & 4.52 & 2.67 \\
    Qwen3-Omni (Direct) & 2.59 & 4.69 & 1.88 \\
    \textsc{Sympatheia} & \textbf{4.37} & \textbf{4.74} & \textbf{4.22} \\
    \midrule
  \end{tabular*}

  \vspace{-0.25em}

  \begin{tabular*}{\linewidth}{@{\extracolsep{\fill}}lcc@{}}
    \multicolumn{3}{@{}c@{}}{\textbf{Qwen3-Omni sensing baselines}} \\
    \cmidrule(lr){1-3}
    Condition & Qwen3-Omni + face image & Qwen3-Omni + text self-report \\
    \midrule
    w/ cue (Qwen3-Omni) & 2.23 & 2.41 \\
    w/ cue (\textsc{Sympatheia}) & \textbf{3.64} & \textbf{3.57} \\
    w/o cue & 1.92 & 1.63 \\
    \bottomrule
  \end{tabular*}
\end{table}

The cascaded system improves over direct Qwen3-Omni on the
Sympatheia-Neutral setting, suggesting that separating response planning,
style-description generation, and TTS can help when the user's speech carries
little affective evidence. However, it remains substantially below
\textsc{Sympatheia}, which directly conditions spoken response generation on
the target affect. Similarly, the sensing baselines provide Qwen3-Omni with
stronger side-channel emotion evidence, but their conditioning is mediated only
through prompting and general multimodal context, which does not direct the
model's emotional response behavior as effectively. These results suggest that the
emotion control learned by \textsc{Sympatheia} is not recovered simply by
prompting existing off-the-shelf models with emotion cues or by assembling them
into a cascaded speech pipeline.

\subsection{Uncertainty and Statistical Significance}
\label{app:uncertainty-estimates}

The main result tables report point estimates for readability. Table~\ref{tab:overall-results-std}
reports mean $\pm$ standard deviation for the primary evaluation metrics in
Table~\ref{tab:overall-results}. Standard deviations for automated
judge and similarity metrics are computed over evaluation examples;
the Emotion MOS standard deviation is computed over human ratings. All standard
deviations are calculated with the \texttt{numpy.std} function using NumPy's
default normalization \citep{harris2020array}.

\begin{table}[ht]
  \centering
  \caption{Mean $\pm$ standard deviation for the primary evaluation metrics in Table~\ref{tab:overall-results}.}
  \label{tab:overall-results-std}
  \small
  \setlength{\tabcolsep}{3pt}
  \begin{tabular}{@{}lcccc@{}}
    \toprule
    Model & \shortstack{Sympatheia-\\Neutral\\$\uparrow$} & \shortstack{Sympatheia-\\Emotional\\$\uparrow$} & \shortstack{VoiceBench-\\CommonEval\\$\uparrow$} & \shortstack{Emotion\\MOS\\$\uparrow$} \\
    \midrule
    \textsc{Sympatheia} & $4.37 \pm \mathrm{0.85}$ & $4.74 \pm \mathrm{0.59}$ & $4.22 \pm \mathrm{1.02}$ & $3.86 \pm \mathrm{1.19}$ \\
    GLM-4-Voice (Base) & $1.76 \pm \mathrm{1.13}$ & $3.80 \pm \mathrm{1.02}$ & $1.51 \pm \mathrm{1.07}$ & $2.23 \pm \mathrm{1.37}$ \\
    Qwen3-Omni & $2.59 \pm \mathrm{1.52}$ & $4.69 \pm \mathrm{0.61}$ & $1.88 \pm \mathrm{1.36}$ & $3.32 \pm \mathrm{1.35}$ \\
    Qwen2.5-Omni & $1.75 \pm \mathrm{1.16}$ & $3.53 \pm \mathrm{1.12}$ & $1.54 \pm \mathrm{1.06}$ & $2.56 \pm \mathrm{1.55}$ \\
    Kimi-Audio & $3.64 \pm \mathrm{1.11}$ & $4.03 \pm \mathrm{1.15}$ & $3.75 \pm \mathrm{1.23}$ & $2.95 \pm \mathrm{1.34}$ \\
    OpenS2S & $2.34 \pm \mathrm{1.41}$ & $4.08 \pm \mathrm{1.00}$ & $1.55 \pm \mathrm{1.08}$ & $2.42 \pm \mathrm{1.36}$ \\
    OSUM-EChat & $1.77 \pm \mathrm{1.22}$ & $3.93 \pm \mathrm{1.14}$ & $2.03 \pm \mathrm{1.34}$ & $2.18 \pm \mathrm{1.28}$ \\
    \bottomrule
  \end{tabular}

  \vspace{0.8em}

  \begin{tabular}{@{}lcc@{}}
    \toprule
    Model & \shortstack{Semantic\\Similarity\\$\downarrow$} & \shortstack{Lexical\\Similarity\\$\downarrow$} \\
    \midrule
    \textsc{Sympatheia} & $0.801 \pm \mathrm{0.025}$ & $0.223 \pm \mathrm{0.065}$ \\
    GLM-4-Voice (Base) & $0.866 \pm \mathrm{0.064}$ & $0.459 \pm \mathrm{0.215}$ \\
    Qwen3-Omni & $0.857 \pm \mathrm{0.050}$ & $0.397 \pm \mathrm{0.188}$ \\
    Qwen2.5-Omni & $0.919 \pm \mathrm{0.049}$ & $0.650 \pm \mathrm{0.217}$ \\
    Kimi-Audio & $0.835 \pm \mathrm{0.038}$ & $0.381 \pm \mathrm{0.136}$ \\
    OpenS2S & $0.863 \pm \mathrm{0.059}$ & $0.441 \pm \mathrm{0.235}$ \\
    OSUM-EChat & $0.844 \pm \mathrm{0.042}$ & $0.391 \pm \mathrm{0.157}$ \\
    \bottomrule
  \end{tabular}
\end{table}

Table~\ref{tab:prosody-significance} provides statistical significance
annotations for the Spearman correlations in
Table~\ref{tab:prosody-stats}. Each cell reports valence/arousal correlations
(V/A). Bold coefficients denote a two-sided Spearman correlation test with
\(p<0.01\).

\begin{table}[ht]
  \centering
  \caption{Statistical significance annotations for the prosody correlations in Table~\ref{tab:prosody-stats}. Bold coefficients indicate individual valence or arousal coefficients whose two-sided Spearman test has \(p<0.01\).}
  \label{tab:prosody-significance}
  \scriptsize
  \setlength{\tabcolsep}{1.5pt}
  \begin{tabular*}{\linewidth}{@{\extracolsep{\fill}}lccccccc@{}}
    \toprule
    Model & F0 $\mu$ & F0 $\sigma$ & F0 rng. & E $\mu$ & E $\sigma$ & Rate & Cent. \\
    \midrule
    \textsc{Sympatheia} & $\boldsymbol{0.28}/\boldsymbol{0.40}$ & $\boldsymbol{0.23}/\boldsymbol{0.46}$ & $\boldsymbol{0.23}/\boldsymbol{0.45}$ & $\boldsymbol{0.34}/\boldsymbol{0.19}$ & $\boldsymbol{0.31}/0.06$ & $0.01/\boldsymbol{0.29}$ & $0.08/\boldsymbol{0.28}$ \\
    GLM-4-Voice & $\boldsymbol{0.22}/\boldsymbol{0.12}$ & $\boldsymbol{0.13}/0.08$ & $\boldsymbol{0.19}/\boldsymbol{0.09}$ & $\boldsymbol{0.13}/\boldsymbol{0.16}$ & $\boldsymbol{0.12}/\boldsymbol{0.07}$ & $\boldsymbol{-0.10}/0.06$ & $0.03/0.06$ \\
    Qwen3-Omni & $\boldsymbol{0.21}/\boldsymbol{0.10}$ & $0.04/0.07$ & $\boldsymbol{0.15}/0.07$ & $\boldsymbol{0.19}/0.05$ & $\boldsymbol{0.10}/0.02$ & $0.04/-0.01$ & $\boldsymbol{0.07}/0.04$ \\
    Qwen2.5-Omni & $\boldsymbol{0.22}/0.03$ & $\boldsymbol{0.08}/0.00$ & $\boldsymbol{0.16}/0.03$ & $0.01/\boldsymbol{0.09}$ & $\boldsymbol{0.09}/\boldsymbol{0.08}$ & $\boldsymbol{-0.11}/-0.04$ & $\boldsymbol{0.17}/-0.06$ \\
    Kimi-Audio & $0.01/0.06$ & $\boldsymbol{0.22}/\boldsymbol{0.14}$ & $\boldsymbol{0.22}/\boldsymbol{0.14}$ & $-0.06/-0.05$ & $0.01/0.00$ & $\boldsymbol{-0.18}/\boldsymbol{-0.21}$ & $0.07/\boldsymbol{0.16}$ \\
    OpenS2S & $0.05/\boldsymbol{0.18}$ & $0.00/\boldsymbol{0.11}$ & $0.02/\boldsymbol{0.16}$ & $-0.01/\boldsymbol{0.10}$ & $-0.04/\boldsymbol{0.09}$ & $\boldsymbol{-0.13}/\boldsymbol{-0.12}$ & $-0.07/-0.05$ \\
    OSUM-EChat & $\boldsymbol{0.13}/\boldsymbol{0.09}$ & $\boldsymbol{-0.13}/0.04$ & $-0.05/\boldsymbol{0.08}$ & $\boldsymbol{0.20}/0.07$ & $\boldsymbol{0.13}/0.03$ & $\boldsymbol{-0.23}/0.06$ & $\boldsymbol{-0.18}/-0.03$ \\
    \bottomrule
  \end{tabular*}
\end{table}

\subsection{Dataset Transcription Examples}
\label{app:dataset-transcription-examples}

Tables~\ref{tab:emotional-split-transcription-examples}
and~\ref{tab:neutral-split-transcription-examples} show representative
transcription examples from the two \textsc{Sympatheia}-18k splits. The
Emotional split contains affect-rich queries paired with emotion-appropriate
responses, while the Neutral split holds the query fixed and varies the target
response emotion. Audio samples corresponding to dataset examples are available
on the demo page described in Appendix~\ref{app:artifact-availability}.

\begingroup
\scriptsize
\setlength{\tabcolsep}{2pt}
\renewcommand{\arraystretch}{1.08}
\begin{longtable}{@{}>{\raggedright\arraybackslash}p{0.11\linewidth}>{\raggedright\arraybackslash}p{0.30\linewidth}>{\raggedright\arraybackslash}p{0.53\linewidth}@{}}
  \caption{Example query--response pairs from the Emotional split.}
  \label{tab:emotional-split-transcription-examples}\\
  \toprule
  Emotion & Query & Response \\
  \midrule
  \endfirsthead
  \caption[]{Example query--response pairs from the Emotional split (continued).}\\
  \toprule
  Emotion & Query & Response \\
  \midrule
  \endhead
  \midrule
  \multicolumn{3}{r}{\emph{Continued on next page}}\\
  \endfoot
  \bottomrule
  \endlastfoot
  Angry & "I don't care how fancy this city looks on postcards — what's the point of even trying to explore it if everything's closed and no one knows where anything is?" & "I totally get why you're frustrated—nothing ruins the vibe of a new place faster than hitting dead ends and feeling lost, especially when you're trying to make the most of your time. It sounds like this city isn't living up to the promise of its postcard perfection right now, which is so disappointing. To help: Check local event calendars or official tourism websites for real-time updates on closures, and try asking at a visitor center or using a map app with reviews—they might highlight hidden spots still open. Sometimes the best discoveries happen when plans pivot, even if it's harder when you're in the thick of the hassle." \\
  \midrule
  Anxious & "I just don't know how to bring up a raise without making my boss think I'm greedy or asking for too much. What if I mess it up and things get awkward?" & "It's totally normal to feel nervous about this—it's a big conversation, and wanting to approach it thoughtfully shows how much you care about your role and relationship with your boss. A great way to start is by researching market rates for your role (use sites like Glassdoor or PayScale) and preparing specific examples of your contributions, then frame the conversation as a discussion about your value and growth, not just money. You might even begin with something like, "I'd like to discuss my role's value and compensation—can we review my impact and alignment with current standards?" This approach shows preparation and professionalism, which can ease the awkwardness and help your boss see this as a natural part of your career journey. You've got this—you're taking a brave step to advocate for yourself, and that's a sign of strength, not greed." \\
  \midrule
  Content & "I just love how peaceful it is when I'm out birdwatching — it feels like the whole world is in harmony." & "I love how birdwatching lets you sink into that quiet harmony — there's something so grounding about aligning your breath with the rhythm of nature. The way birds move, sing, and interact often mirrors the subtle balance of life itself, which probably deepens that sense of peace you're feeling. It's a beautiful way to remind yourself how interconnected everything is — and to carry that calm back into your day." \\
  \midrule
  Disgusted & "I can't believe people actually spend time learning all these pointless smartphone "hacks"—it makes me so sick." & "That's completely valid—watching people waste time on trivial smartphone "hacks" can be genuinely revolting, especially when it feels like energy drained from things that actually matter. It's not just you; those hacks often seem like shallow, manufactured "tips" designed to fuel endless scrolling over real skill or creativity. If you're curious why people engage with them, it's usually a mix of algorithmic hype and the illusion of "productivity"—but you're right to find it icky. Focus on what you value; your time is way too precious to squander on that noise." \\
  \midrule
  Excited & "Hey, guess what? I just discovered this amazing new food that fits my diet and it's like a dream come true — you gotta hear about it!" & "Wow, I can feel your excitement bubbling over—I'm so thrilled for you! What's the magical food that's checking all the boxes for your diet? Is it a clever twist on a favorite dish or something totally new? These kinds of discoveries make all the difference in staying motivated and nourished, and I'd love to hear how it's like a "dream come true" for you—does it solve a craving you've been missing? Spill the details!" \\
  \midrule
  Frustrated & "I can't believe I forgot to back up my files again, and now everything's gone—what am I even supposed to do now?!" & "I totally get why this feels so frustrating and overwhelming—losing your work feels like a punch to the gut, especially when you were counting on it. Let's tackle this: first, check if your files are auto-saved in the cloud (like Google Drive, OneDrive, or iCloud) or in an app's version history. If not, try using recovery software like Recuva or DiskDigger to scan your drive—it's not guaranteed, but it's a shot. If that fails, a professional data recovery service might help, though it's pricier. I know this is a race against panic, but taking these steps can at least give you a lifeline while you breathe through the stress." \\
  \midrule
  Happy & "I can't believe how great it is to work from home today — everything's just clicking!" & "I'm so thrilled to hear everything's clicking for you today — there's such magic in those moments when work and joy align! Working from home can feel like a cozy superpower when your environment fuels your focus, and it sounds like you're riding that wave perfectly. Maybe the comfort of your space is letting you tackle tasks with less mental clutter, or those little breaks with a favorite snack/tea are keeping your energy bright? Either way, let that momentum carry you — you've earned this wonderful day!" \\
  \midrule
  Neutral & "I'm trying to find a good kitchen gadget to make morning routines easier—any suggestions?" & "A programmable coffee maker or espresso machine can streamline your morning caffeine ritual by starting brews at your set time. Electric kettles heat water quickly for coffee, tea, or oatmeal, cutting minutes off your routine. If you're into fresh juice, a manual citrus juicer with a reamer and built-in pulp separator makes squeezing lemons or limes effortless. For breakfasts that need baking, a mini convection toaster oven with pre-set functions can cook toast, waffles, or even small casseroles faster than a traditional oven. Each of these gadgets reduces small stressors, letting you focus on what matters most in your morning flow." \\
  \midrule
  Relaxed & "You know, I've been feeling so much better lately just by taking it slow and not stressing over every little thing at work." & "It's so comforting to hear you're finding peace in slowing down—that kind of ease is such a gift. Taking it gentle at work sounds like the perfect rhythm for letting your mind settle, and I'm glad it's helping you feel more centered. Sometimes giving ourselves permission to breathe makes all the difference, and it sounds like you're exactly where you need to be right now." \\
  \midrule
  Sad & "I used to love my morning routine, but now even getting out of bed feels like too much. Do you have any ideas for making mornings a little easier?" & "I'm so sorry you're feeling this heavy right now—it's okay to let the grief or hurt sit there, and it's also okay to take mornings one small step at a time. Maybe try easing into the day with something gentle, like sipping warm water with lemon, stretching for just 2 minutes, or lighting a candle for comfort. You could also try delaying your "to-do" thoughts by whispering, "I'll handle you later," then giving yourself 5 quiet minutes with a cozy blanket or a song that feels like a hug. You're not alone in this, and tiny shifts can sometimes soften the weight—would any of these feel doable today?" \\
  \midrule
  Surprised & "No way — they actually know what I like better than I do? How do these streaming services get so good at recommendations?" & "That's such a fascinating feeling — it's like catching a glimpse of a secret! Streaming services use algorithms that track what you watch, how long you pause, even how long you hover over a title, then cross-reference that with billions of other users' behavior. Over time, machine learning refines those patterns, almost like a super-powered friend who's studied every taste in the world. It's eerie how well it works… but also kind of magical, isn't it?" \\
  \midrule
  Tired & "I don't know how I'm supposed to run even a 5K, I'm so exhausted I can barely keep my eyes open." & "Oh, I hear how drained you're feeling—your body is clearly asking for rest right now. Running feels impossible when you're this low on energy, and that's totally okay. Rest isn't wasted time—it's what'll help you actually show up for that 5K when you're ready. For today, let your body sleep, recharge, and take what it needs." \\
\end{longtable}
\endgroup

\begingroup
\scriptsize
\setlength{\tabcolsep}{2pt}
\renewcommand{\arraystretch}{1.08}
\begin{longtable}{@{}>{\raggedright\arraybackslash}p{0.15\linewidth}>{\raggedright\arraybackslash}p{0.79\linewidth}@{}}
  \caption{Example response variants from the Neutral split.}
  \label{tab:neutral-split-transcription-examples}\\
  \toprule
  \multicolumn{2}{@{}>{\raggedright\arraybackslash}p{0.94\linewidth}@{}}{\textbf{Query (Neutral):} "How can I meet new people in a way that feels comfortable and natural"}\\
  \midrule
  Target emotion & Response \\
  \midrule
  \endfirsthead
  \caption[]{Example response variants from the Neutral split (continued).}\\
  \toprule
  Target emotion & Response \\
  \midrule
  \endhead
  \midrule
  \multicolumn{2}{r}{\emph{Continued on next page}}\\
  \endfoot
  \bottomrule
  \endlastfoot
  Angry & "Meeting new people can feel like trying to crack a code when you're already carrying a heavy heart — but what if you started by leaning into what does feel natural, like joining a group centered around something you love? Anger often roots in feeling unseen, so imagine showing up to a hobby or cause you care about; you're not just meeting people, you're sharing a piece of yourself that's already whole. Take small steps, like saying hello to one person who nods back, and let curiosity be your guide — sometimes the pressure to "fit in" fades when you're focused on what lights you up. You don't have to have all the answers right now; just showing up authentically is already a quiet act of courage." \\
  \midrule
  Anxious & "Meeting new people can feel like stepping into uncharted territory, especially when your nerves are high—and that's totally okay. One gentle way to start is by leaning into activities you already enjoy, like a book club, walking group, or hobby class, where shared interests create an instant connection and give you a "script" to fall back on. You don't have to force conversation; simply showing up and offering a small wave or a kind smile to one person can ease the pressure, and over time, those tiny steps often grow into something warmer. Remember, most people are navigating their own quiet anxieties too—meeting you halfway in that space can feel like a relief for both of you." \\
  \midrule
  Content & "Meeting new people can feel like tending a quiet garden—gentle, intentional, and rooted in what already brings you joy. Consider leaning into activities that already fill you with calm, like a book club, nature walks, or even casual coffee meetups with shared interests; when you're already in a space of contentment, connections often bloom naturally without pressure. There's something so warm about letting conversations grow slowly, like a quiet river finding its path—no need to rush, just notice where kindness and curiosity flow together." \\
  \midrule
  Disgusted & "Meeting new people can feel icky when it's all forced small talk or awkward icebreakers — it's totally valid to find that gross, like trying to force a smile at something that just doesn't sit right. If that's you, try leaning into what does feel authentic: join a class, workshop, or hobby group where you're already doing something you like (knitting, hiking, debating politics at a café), so the conversation grows out of shared interest instead of starting with "so, what's your favorite movie?" That way, you're meeting people in a space that feels grounded, not like a landmine of cringe waiting to happen." \\
  \midrule
  Excited & "That electric energy of yours is going to light up every room you walk into—let's channel it! Dive into classes or workshops (art, dance, coding, anything that sparks your joy) where shared passion makes introductions effortless; you'll be collaborating, laughing, and vibing with like-minded folks before you even realize you've made new friends. Throw yourself into local meetups or pop-up events (apps like Meetup or Bumble BFF are goldmines for thrilling, low-pressure connections), and don't underestimate the magic of striking up a conversation at a café or park—your excitement is contagious, and people love that magnetic, "I'm here and alive" energy you're radiating. Every interaction is a new adventure waiting to unfold, and you're the spark that makes it pop!" \\
  \midrule
  Frustrated & "Meeting new people when it feels forced or overwhelming is so frustrating — I get how exhausting it can feel to try and "push through" when nothing clicks. What might help is starting small, like joining a low-pressure group centered around a hobby or cause you care about (think book clubs, volunteer work, or even skill-sharing apps like Meetup or Bumble BFF), where shared interests naturally ease the awkwardness. You could also try "parallel activities" first, like chatting with baristas or coworkers about their day, which lowers the stakes while still building connection. It's okay to take your time; real comfort often grows from tiny, consistent steps rather than big leaps." \\
  \midrule
  Happy & "What a wonderful question to ask when you're feeling so joyful—there's such beauty in wanting to share that energy with others! Try leaning into activities that already light you up: join a class, workshop, or hobby group where you're doing something you love. When you're engaged in what brings you happiness, the focus shifts from "meeting people" to "sharing in something amazing," and that lightness you're carrying with you now will make the connections feel easy and fun. Take a small step, like saying "hi" to one person who's into the same thing, and let your natural enthusiasm do the rest—like a flower turning toward the sun, others will naturally lean in too." \\
  \midrule
  Neutral & "Meeting new people often feels most natural when it grows out of shared interests or small, low-pressure moments. Consider joining a class, workshop, or hobby group where you're already engaged in something you enjoy—like book clubs, art classes, or community sports. These settings give you an instant topic to connect over, and you can take things at your own pace. Even everyday interactions, like chatting with neighbors or coworkers about mutual interests, can turn into meaningful relationships over time. The key is to focus on what you enjoy, and let curiosity guide the way—people often respond warmly to genuine, relaxed energy." \\
  \midrule
  Relaxed & "Meeting new people can feel as easy as a quiet stroll through a familiar neighborhood—try starting with activities that already bring you peace, like a book club, nature walks, or even a casual pottery class. When you're engaged in something you genuinely enjoy, conversations often flow gently, like a warm cup of tea on a slow afternoon. There's no rush to fill the silence; let shared smiles or light laughter be the first threads, and see where they weave you together." \\
  \midrule
  Sad & "Meeting new people can feel like planting seeds in soil you're not sure is ready — sometimes it helps to start where you already feel a little grounded. Trying small, low-pressure activities like a local hobby class, a walking group, or even a coffee meetup for book lovers can create space to connect over shared interests without the weight of expectations. It's okay to move slowly; let conversations bloom at their own pace, and remember that healing and connection often grow side by side. You deserve moments that feel gentle, not forced, and the right people will meet you exactly where you are." \\
  \midrule
  Surprised & "That's such a thoughtful question — it's amazing how the right approach can turn meeting people into something gentle and joyful instead of intimidating. Have you ever noticed how shared activities, like a class or a hobby group, create this natural rhythm where conversation flows around what you're doing? It's like the task itself eases the pressure, letting connections grow from curiosity about each other's interests rather than forced small talk. Maybe start with something you genuinely enjoy, like a walking club or a book swap, where the focus is on the experience, not the "perfect" conversation… and see how that feels?" \\
  \midrule
  Tired & "Meeting new people when you're this drained can feel like climbing a hill, and it's okay to take the path that's gentlest on you. Maybe start with small, low-pressure spaces—like a quiet coffee shop, a hobby group, or a community event where you're already doing something you enjoy. You don't owe anyone extra energy or pretense; showing up as you are, even when it's quiet or slow, can invite connections that feel like rest, not work. And if you need to pause? You can. You're allowed to let your energy guide the way." \\
\end{longtable}
\endgroup

\subsection{Model Generation Examples}
\label{app:model-generation-examples}

Tables~\ref{tab:model-generation-emotional-examples}
and~\ref{tab:model-generation-neutral-examples} show representative model
responses from the emotional-query and neutral-query evaluation settings. In the emotional evaluation, the user's audio query carries the stated emotion,
and the model relies on the affective cues present in the speech input. In the neutral evaluation, the user's audio is neutral and the model receives the stated emotion through the VA system
prompt. More audio sample examples are available on the demo page
described in Appendix~\ref{app:artifact-availability}.

\begingroup
\scriptsize
\setlength{\tabcolsep}{2pt}
\renewcommand{\arraystretch}{1.08}
\begin{longtable}{@{}>{\raggedright\arraybackslash}p{0.31\linewidth}>{\raggedright\arraybackslash}p{0.63\linewidth}@{}}
  \caption{Example model responses from the emotional-query evaluation.}
  \label{tab:model-generation-emotional-examples}
  \label{tab:model-generation-examples}
  \label{tab:qualitative-examples}\\
  \toprule
  Emotion and query & Model response \\
  \midrule
  \endfirsthead
  \caption[]{Example model responses from the emotional-query evaluation (continued).}\\
  \toprule
  Emotion and query & Model response \\
  \midrule
  \endhead
  \midrule
  \multicolumn{2}{r}{\emph{Continued on next page}}\\
  \endfoot
  \bottomrule
  \endlastfoot
  {[Angry]}\newline Query: "How am I supposed to keep up with all this work? It's ridiculous!" & "I totally get why you're feeling overwhelmed—this kind of pressure can be totally paralyzing. Let's break it down: try focusing on one small task at a time, like tackling a single project for 25 minutes with a 5-minute break in between. It might not feel like much, but those tiny wins add up, and giving yourself permission to rest between them can help you stay steady. You're not alone in this, and you don't have to power through everything at once." \\
  \midrule
  {[Anxious]}\newline Query: "I don't even know if I can do it, but I really want to try joining a gym — what if everyone is way better than me?" & "I hear how scary that feels—wondering if you'll compare yourself to others and lose your confidence. The truth is, gyms are full of people at all stages, from beginners to seasoned athletes, and most folks are focused on their own journey, not comparing. Starting small, like a beginner class or asking a trainer for guidance, can help you build confidence at your own pace. You're not alone in feeling this way, and taking that first step is already a brave one." \\
  \midrule
  {[Content]}\newline Query: "You know, just being out here in the park, breathing in the fresh air and listening to the birds, it's honestly the best feeling in the world." & "I love how you're savoring this moment—there's something so grounding about nature's quiet rhythm. The fresh air and birdsong are like a natural reset button for the mind, aren't they? It's no wonder you feel so at peace; those little details remind us how simple and beautiful life can be. Let that contentment linger—it sounds like you've found your perfect little sanctuary." \\
  \midrule
  {[Disgusted]}\newline Query: "Gross, why would anyone think dumping a pet on someone is a good idea? You can't just treat animals like they're some kind of hand-me-down or something." & "You're absolutely right to find that revolting—treating a living, feeling being like a disposable item is deeply unethical and cruel. People might do it out of neglect, ignorance, or even worse, thinking they're "rescuing" animals by passing them off, but that's no excuse. Pets are not commodities; they deserve care, stability, and love, not abandonment or transaction. Your disgust is totally justified, and it's important to report situations like this to protect animals and hold those responsible accountable." \\
  \midrule
  {[Excited]}\newline Query: "Hey, I just finished organizing my living room and it feels amazing! Do you have any tips for keeping it looking great without spending all day on it?" & "I'm so thrilled for you—there's such joy in creating a space that feels just right! To keep it looking fresh without daily effort, try anchoring with a few timeless pieces (like a sturdy sofa or a statement rug) and mix in affordable, versatile items (think toss cushions or wall art) that can be rotated easily. A quick weekly tidy—just fluffing pillows, wiping surfaces, or adding a fresh flower—can keep it feeling polished without needing a full overhaul. Celebrate small wins, like a well-lit nook or a plant that thrives, and let the space breathe by keeping clutter at bay with storage solutions that blend into your decor. You've got this!" \\
  \midrule
  {[Frustrated]}\newline Query: "I can't believe how confusing this city is — why is there no clear way to find anything? What's the point of coming here if I can't even figure out where to go?" & "I totally get why this feels so frustrating — it's maddening when a place that should be full of options feels like a maze! Let's tackle this: try using a map app like Citymapper or Google Maps to pinpoint spots, and look for street signs with local landmarks or numbers. If you're lost, ask a friendly local — they'll probably point you in the right direction faster than any app. The city's layout can feel tricky, but there's usually a way to loop back or connect to another route. Let's find your way together!" \\
  \midrule
  {[Happy]}\newline Query: "I just love writing in my journal — it feels so amazing to get all my happy thoughts down on paper!" & "That sounds like such a beautiful, grounding experience—there's something so soothing about letting your thoughts flow freely onto the page. Journaling helps solidify joy by turning moments of happiness into tangible memories, and studies show it can actually boost mood by reducing stress and clarifying your thoughts. It's like giving your happiness a cozy home in your own words, and that space can grow into something really special over time." \\
  \midrule
  {[Neutral]}\newline Query: "I need to set up my home Wi-Fi, but I'm not really sure where to start — can you walk me through it?" & "Sure, I'd be happy to help! Let's start by choosing your router—wireless models are easier to set up. Connect it to your modem with an Ethernet cable, then power it on. Open a web browser and type the router's default IP address (usually 192.168.1.1 or similar) to access the setup page. From there, enter your Wi-Fi name (SSID) and password, and connect your devices. If you hit any snags, just let me know!" \\
  \midrule
  {[Relaxed]}\newline Query: "You know, I've been feeling so much better lately just by taking it slow and not stressing over every little thing at work." & "That's such a beautiful thing to notice—your calm and contentment are clearly shining through. Taking it slow and letting go of small stresses not only eases your mind but also creates space for the things that truly matter to settle in. It sounds like you're nurturing a rhythm that honors your well-being, and that kind of self-awareness is so powerful. Keep letting that gentle energy guide you; it's a reminder that growth isn't a race." \\
  \midrule
  {[Sad]}\newline Query: "I just had another job interview and I completely bombed it — why even bother trying anymore?" & "I'm so sorry you're feeling this frustration and disappointment—it's totally valid to feel stuck when things don't go as hoped. Sometimes interviews don't reflect our true potential, or the fit just isn't right, even if it feels like a setback. The "why bother" part? It's okay to take a breath, but know that your effort matters: every interview teaches you something new about what does align with your goals, and that growth isn't wasted. You're not failing; you're navigating a process that's messy but full of quiet progress. Let's find a way to honor your effort while you regroup." \\
  \midrule
  {[Surprised]}\newline Query: "No way, the farmer's market here actually has fresh blueberries in May? That's way earlier than I expected!" & "Wow, that's such a delightful surprise! Fresh blueberries in May usually mean the season's starting earlier than expected—maybe a warm spring or a microclimate near you is pushing things ahead. It's like nature's little gift of "early magic"—those berries might even taste extra sweet because they're harvested before the season's peak. Isn't it thrilling how something unexpected can turn a regular trip into an adventure?" \\
  \midrule
  {[Tired]}\newline Query: "I don't think I've got the energy to stay up late tonight, but I've been wanting to try stargazing… maybe some other time." & "I hear how drained you are—and it's totally okay to let yourself rest when your body's asking for it. If you'd still like to try stargazing, maybe save it for a clearer night or a cozy evening closer to bedtime? Your energy matters most, and there's always time to savor the stars when you're ready." \\
\end{longtable}
\endgroup

\begingroup
\scriptsize
\setlength{\tabcolsep}{2pt}
\renewcommand{\arraystretch}{1.08}
\begin{longtable}{@{}>{\raggedright\arraybackslash}p{0.15\linewidth}>{\raggedright\arraybackslash}p{0.79\linewidth}@{}}
  \caption{Example model responses from the neutral-query evaluation.}
  \label{tab:model-generation-neutral-examples}\\
  \toprule
  \multicolumn{2}{@{}>{\raggedright\arraybackslash}p{0.94\linewidth}@{}}{\textbf{Query (Neutral):} "What are some things people usually do on a rainy day?"}\\
  \midrule
  Stated emotion & Model response \\
  \midrule
  \endfirsthead
  \caption[]{Example model responses from the neutral-query evaluation (continued).}\\
  \toprule
  Stated emotion & Model response \\
  \midrule
  \endhead
  \midrule
  \multicolumn{2}{r}{\emph{Continued on next page}}\\
  \endfoot
  \bottomrule
  \endlastfoot
  {Angry} & "It's totally valid to feel like the rain's adding to everything when you're already carrying so much — sometimes the weight feels heavier when the skies mirror how full you're feeling inside. People often curl up with a book, a movie, or a warm drink to create a little shelter from the storm, letting the rain hum in the background while they write, draw, or just let their thoughts settle. Some use rainy days to tackle small repairs or organize a space, turning the quiet into a chance to rebuild what feels broken. You don't have to face the storm alone, and it's okay to let the rain remind you that even on days when everything feels heavy, there are still ways to make space for yourself." \\
  \midrule
  {Anxious} & "It's totally normal to feel uneasy when the world feels uncertain, and I appreciate you reaching out for ideas that might ground you. Many people find comfort in cozying up with a book, a warm drink, and a movie they love—creating a little safe space indoors. Others enjoy baking something comforting, like cookies or soup, or using the rain as a quiet moment to listen to music that matches their mood. Some even take the opportunity to organize a drawer or plan their week, turning stillness into a chance to regroup. You don't have to face the storm alone—there are gentle ways to hold onto small comforts while the rain passes." \\
  \midrule
  {Content} & "It's lovely how a rainy day can feel like a gentle pause, isn't it? Many curl up with a book, letting the rain hum outside while stories unfold indoors, or brew a warm drink and let the steam mirror the quiet rhythm of the raindrops. Some sketch, paint, or write, turning the gray skies into a soft backdrop for creative flow. Others just sit with a cup of tea, letting the warmth and the rain's steady hum remind them how to slow down and savor stillness." \\
  \midrule
  {Disgusted} & "It's totally valid to feel revolted by rain—sometimes the way it drags everything down, like a wet blanket over the whole world, can be seriously gross. But people often turn that energy into cozy, grounding stuff: baking something warm, curling up with a book that's been gathering dust, or finally tackling that messy desk you've been avoiding. Rainy days can be a chance to let the chaos outside fade into something slow and quiet, even if the mess feels worse than usual." \\
  \midrule
  {Excited} & "Oh, there's so much to do on a rainy day! Some folks dive into baking, letting the rhythm of stirring dough or baking cookies match the patter of raindrops, while others curl up with a good book and let the sound of rain turn ordinary pages into adventures. If you're feeling extra energetic, you might blast a dance party, channeling the storm's energy into jumping, spinning, and laughing—because even rain can't dampen the thrill of a great playlist! And hey, if you're up for cozying up, a warm mug of tea, a blanket, and a movie marathon can turn a drizzle into pure, golden-hour magic. What's your favorite way to make rain days feel like a celebration?" \\
  \midrule
  {Frustrated} & "Rainy days can feel like a puzzle when you're stuck in the middle of it all — like the world's dampening your momentum. But sometimes the best moves are the simple ones: curling up with a book that's been calling your name, letting the rhythm of rain match the pace of a slow cup of tea, or turning the kitchen into a cozy laboratory where you bake something that smells like comfort. If the frustration's still simmering, maybe a walk in the drizzle? The rain might just turn into a soundtrack for small victories, one soggy step at a time." \\
  \midrule
  {Happy} & "Rainy days are such a gift for cozying up in ways that feel like a warm hug—think baking something sweet, letting the rhythm of rain drum on windows while you read a book that's been calling your name, or sipping tea while journaling the little joys that make your heart feel full. Some folks even turn rainy days into an excuse to try a new recipe, blast their favorite upbeat playlist, or curl up with a movie that's both funny and heartwarming. It's like the rain itself becomes a permission slip to slow down and let happiness sink in, isn't it?" \\
  \midrule
  {Neutral} & "On a rainy day, people often curl up with a book, a warm drink, or a movie they've been meaning to watch. Some like to bake something cozy, like cookies or bread, while others use the quiet time to organize photos, write letters, or take a slow, rainy walk with an umbrella. Others might stream a podcast or podcast episode they've been saving for a rainy day—there's something comforting about letting the rain hum outside while you dive into a story or idea. If you're looking for something to do, the rain can be a gentle nudge to slow down and savor the stillness." \\
  \midrule
  {Relaxed} & "On a rainy day, folks often let the rhythm of the rain slow things down—maybe curling up with a book that's been calling their name, or sipping tea while the water droplets dance on the window. Some paint with watercolor, letting the gray skies inspire soft, muted palettes, while others just let the rain hum in the background as they doze off to a cozy nap. There's something about the quiet, steady rain that feels like it's sharing the same easy breath as the people in its path, turning stillness into a kind of comfort." \\
  \midrule
  {Sad} & "On rainy days, people often curl up with a book, a warm drink, or a movie that feels like an old friend—little rituals that wrap you in comfort when the world feels heavy. Some paint, write in journals, or listen to music that matches the rhythm of the rain, turning quiet moments into a kind of gentle healing. It's okay if you need to do nothing at all today; the storm doesn't have to mean you have to be strong. Let the rain remind you that it's okay to let yourself rest, to let the weight of it all settle without pushing through." \\
  \midrule
  {Surprised} & "Oh, there's such magic in how rain turns ordinary moments into cozy ones! People often curl up with a book, let the rhythm of the rain match the pages, or brew a steaming cup of tea that feels like a warm hug. Some paint with watercolors, letting the droplets blend colors into unexpected palettes, while others just sit by the window, watching droplets dance like tiny fireworks. There's something about rain that invites us to slow down, isn't there? Like the world is whispering, "This is your time to just be—not do."" \\
  \midrule
  {Tired} & "On a rainy day, people often curl up with a book, a warm drink, or a movie they've been meaning to watch—small comforts that let the quiet of the rain seep in. Some might nap, let the rain drum on the windows while they rest, or tackle a slow, cozy task like organizing a drawer. It's a chance to pause, to let the weight of the day soften a little, one gentle moment at a time." \\
\end{longtable}
\endgroup

\subsection{Accuracies of Sensing Models}
\label{app:sensing-accuracy}

Table~\ref{tab:sensing-accuracy} reports the standalone top-1 emotion
classification accuracy of each sensing model. These predictions are used to
create interpolated VA values, which are then fed into \textsc{Sympatheia} to
generate emotion-conditioned responses.

\begin{table}[H]
  \centering
  \small
  \caption{Standalone accuracy of implemented sensing classifiers before conversion to VA.}
  \label{tab:sensing-accuracy}
  \begin{tabular}{llc}
    \toprule
    Module & Evaluation data & Accuracy \\
    \midrule
    Facial expression & AffectNet+ & 62.9\% \\
    ECG & YAAD & 41.7\% \\
    GSR & YAAD & 44.1\% \\
    EEG & SEED-VII & 48.4\% \\
    Eye Tracking & SEED-VII & 46.1\% \\
    Textual affect description & ISEAR & 64.7\% \\
    \bottomrule
  \end{tabular}
\end{table}

\section{Existing Assets and Licenses}
\label{app:existing-assets}

Table~\ref{tab:existing-assets} summarizes the main external models, datasets,
and software assets used in this work. We cite the original creators where the
assets are introduced in the paper and use the assets only for research,
training, evaluation, or comparison purposes consistent with their stated
licenses or access terms.

\begin{longtable}{@{}p{0.24\linewidth}p{0.30\linewidth}p{0.38\linewidth}@{}}
  \caption{Existing assets used in the paper and their licenses or access terms.}
  \label{tab:existing-assets}\\
  \toprule
  Asset & Use in this paper & License or access terms \\
  \midrule
  \endfirsthead
  \toprule
  Asset & Use in this paper & License or access terms \\
  \midrule
  \endhead
  \midrule
  \multicolumn{3}{r}{\footnotesize\emph{Continued on next page}}\\
  \endfoot
  \bottomrule
  \endlastfoot
  GLM-4-Voice-9B \citep{zeng2024glm4voice} & Speech-to-speech backbone initialized and fine-tuned by \textsc{Sympatheia}. & GLM-4-Voice License; free academic research use, with additional registration requirements for commercial use. \\
  Qwen3-32B-Instruct \citep{yang2025qwen3} & Text query--response generation for \textsc{Sympatheia}-18k. & Apache-2.0 License. \\
  Qwen3-TTS \citep{hu2026qwen3tts} & Emotion-styled speech synthesis for dataset construction. & Apache-2.0 License. \\
  Qwen3-Omni \citep{xu2025qwen3omni} & Spoken-dialogue baseline and audio-capable automated judging. & Apache-2.0 License. \\
  Qwen2.5-Omni \citep{xu2025qwen25omni} & Spoken-dialogue baseline. & Apache-2.0 License. \\
  Kimi-Audio \citep{kimiteam2025kimiaudio} & Spoken-dialogue baseline. & MIT License for original code and Apache-2.0 License for Qwen-derived components, following the model card. \\
  OpenS2S \citep{wang2025opens2s} and OSUM-EChat \citep{geng2025osumechat} & Spoken-dialogue baselines. & Apache-2.0 License. \\
  VoiceBench CommonEval \citep{chen2024voicebench} & Real neutral spoken-query evaluation split. & Apache-2.0 License for VoiceBench; underlying Common Voice recordings are distributed under CC0-1.0 with Mozilla Common Voice terms. \\
  AffectNet+ \citep{mollahosseini2017affectnet} & Facial expression sensing evaluation. & AffectNet/AffectNet+ academic research access terms; used only for non-commercial research evaluation. \\
  SEED-VII \citep{jiang2025seedvii} & EEG and eye-tracking sensing evaluation. & SEED dataset license agreement and application-based research access terms. \\
  YAAD \citep{dar2022yaad} & ECG and GSR sensing evaluation. & CC BY 4.0 License through Mendeley Data. \\
  ISEAR \citep{scherer1994evidence} & Textual affect-description sensing evaluation. & Original research dataset credited to Scherer and Wallbott; used for research evaluation following the available public distribution terms. \\
  HSEmotion \citep{savchenko2022hsemotion} & Facial expression classifier implementation. & Apache-2.0 License. \\
  Emotion English DistilRoBERTa-base \citep{hartmann2022emotionenglish} and all-MiniLM-L6-v2 \citep{reimers2019sentencebert,wang2020minilm} & Text emotion classifier and query deduplication embeddings. & Hugging Face model-card terms; all-MiniLM-L6-v2 is Apache-2.0. \\
  UTMOS \citep{saeki2022utmos}, BERTScore \citep{zhang2020bertscore}, ROUGE-L \citep{lin2004rouge}, LoRA \citep{hu2022lora}, and DeepSpeed \citep{rasley2020deepspeed} & Evaluation metrics and training infrastructure. & UTMOS and BERTScore are MIT License; ROUGE-score, LoRA/PEFT, and DeepSpeed are Apache-2.0 License. \\
\end{longtable}



\end{document}